\documentclass[12pt]{article}

\pdfoutput=1
\DeclareFontFamily{T1}{calligra}{}
\DeclareFontShape{T1}{calligra}{m}{n}{<->s*[1.44]callig15}{}
\DeclareMathAlphabet\mathcalligra   {T1}{calligra} {m} {n}
\DeclareMathAlphabet\mathzapf       {T1}{pzc} {mb} {it}
\DeclareMathAlphabet\mathchorus     {T1}{qzc} {m} {n}
\DeclareMathAlphabet\mathrsfso      {U}{rsfso}{m}{n}
\DeclareMathAlphabet\mathfrcal      {T1}{frcursive}{m}{it}
\DeclareFontFamily{T1}{frcursive}{}
\DeclareFontShape{T1}{frcursive}{m}{n}{<->s*[1.44]callig15}{}
\DeclareMathAlphabet\mathfrcal      {T1}{frcursive}{m}{it}

\usepackage{amsmath}
\usepackage{amssymb}
\usepackage{graphicx}
\usepackage[dvipsnames]{xcolor}
\usepackage{soul}
\numberwithin{equation}{section}
\usepackage{array}
\usepackage{mathtools}
\usepackage{dsfont}
\usepackage{mathrsfs}

\usepackage{BOONDOX-uprscr}

\usepackage{tikz}\usetikzlibrary{matrix,fit}
\usepackage{varwidth}
\usepackage{enumerate}
\usepackage{appendix}

\usepackage[margin=1in]{geometry}
\usepackage{nicematrix}

\usepackage[
    backend=bibtex,
    style=alphabetic,
maxbibnames=99,
giveninits=true,
minalphanames=1,
maxalphanames=3,url=false
  ]{biblatex}

\bibliography{SUSYGrassmannian}

\usepackage{mathabx}
\usepackage{empheq}

\usepackage{setspace}

\usepackage{slashed}
\usepackage{upgreek}
\usepackage{appendix}

\usepackage{tabstackengine}

\usepackage{wrapfig}
\usepackage[abs]{overpic}

\usepackage{float}

\fixTABwidth{T}

\newdimen\mytextwidth
\newcommand\rem[2][cyan!40!green]{\noindent\nobreak\hfil\penalty1000\hfilneg
\mytextwidth=\linewidth\advance\mytextwidth by 2mm
\begin{tikzpicture}[baseline=-\the\dimexpr\fontdimen22\textfont2\relax]\node[outer sep=0pt,draw=black,fill=#1,fill opacity=1,text opacity=1,rectangle,rounded corners]{\begin{varwidth}{\mytextwidth}\textcolor{white}{#2}\end{varwidth}};
\end{tikzpicture}\allowbreak
}

\newcommand\whiterem[2][white!]{\noindent\nobreak\hfil\penalty1000\hfilneg
\mytextwidth=\linewidth\advance\mytextwidth by 2mm
\begin{tikzpicture}[baseline=-\the\dimexpr\fontdimen22\textfont2\relax]\node[outer sep=0pt,draw=black,fill=#1,fill opacity=1,text opacity=1,rectangle,rounded corners,line width=1.5pt]{\begin{varwidth}{\mytextwidth}\textcolor{black}{#2}\end{varwidth}};
\end{tikzpicture}\allowbreak
}

\newcommand{\dd}{\partial}

\newcommand{\CP}{\mathsf{CP}}
\newcommand{\CC}{\mathsf{C}}

\renewcommand{\tilde}{\widetilde}

\newcommand{\bea}{\begin{equation}}
\newcommand{\eea}{\end{equation}}
\newcommand{\bear}{\begin{eqnarray}}
\newcommand{\eear}{\end{eqnarray}}
\newcommand{\bearr}{\begin{eqnarray*}}
\newcommand{\eearr}{\end{eqnarray*}}

\renewbibmacro{in:}{}

\usepackage{mdframed}

\newbibmacro{string+doi}[1]{
  \iffieldundef{doi}{#1}{\href{http://dx.doi.org/\thefield{doi}}{#1}}}
\DeclareFieldFormat{title}{\usebibmacro{string+doi}{\mkbibemph{#1}}}
\DeclareFieldFormat[article]{title}{\usebibmacro{string+doi}{\mkbibquote{#1}}}

\newmdenv[
  topline=false,
  bottomline=false,
  rightline=false,
  linewidth=2pt,
  skipabove=\topsep,
  skipbelow=\topsep
]{siderules}

\newmdenv[
  topline=false,
  bottomline=false,
  linewidth=2pt,
  skipabove=\topsep,
  skipbelow=\topsep
]{siderulesright}

\makeatletter
\renewcommand{\@seccntformat}[1]{\csname the#1\endcsname.\quad}
\makeatother

\usepackage{setspace}
\usepackage{xpatch}

\makeatletter
\renewcommand{\@chap@pppage}{
  \clear@ppage
  \thispagestyle{plain}
  \if@twocolumn\onecolumn\@tempswatrue\else\@tempswafalse\fi
  \null\vfil
  \markboth{}{}
  {\centering
   \interlinepenalty \@M
   \normalfont
   \MakeUppercase \appendixpagename\par}
  \if@dotoc@pp
    \addappheadtotoc
  \fi
  \vfil\newpage
  \if@twoside
    \if@openright
      \null
      \thispagestyle{empty}
      \newpage
    \fi
  \fi
  \if@tempswa
    \twocolumn
  \fi
}
\makeatother

\definecolor{navycol}{RGB}{100,150,160}
   \definecolor{pinkcol}{RGB}{242,55,55}
   \definecolor{greencol}{RGB}{50,205,50}

   \definecolor{bluecol}{RGB}{30,144,255}

\usepackage{titlesec}

\titleformat*{\section}{\large\bfseries}
\titleformat*{\subsection}{\normalsize\bfseries}
\titleformat*{\subsubsection}{\normalsize\bfseries}
\titleformat*{\paragraph}{\large\bfseries}
\titleformat*{\subparagraph}{\large\bfseries}
\titlespacing{\author}{-5pt}{-5pt}{-5pt}[-5pt]

\makeatletter
\renewcommand\subsubsection{\@startsection{subsubsection}{3}{\z@}
                                     {-3.25ex\@plus -1ex \@minus -.2ex}
                                     {-1.5ex \@plus -.2ex}
                                     {\normalfont\normalsize\bfseries}}
\renewcommand\subsection{\@startsection{subsection}{3}{\z@}
                                     {-3.25ex\@plus -1ex \@minus -.2ex}
                                     {-1.5ex \@plus -.2ex}
                                     {\normalfont\normalsize\bfseries}}                                     
\makeatother

\DeclareFontFamily{U}{solomos}{}
\DeclareErrorFont{U}{solomos}{m}{n}{10}
\DeclareFontShape{U}{solomos}{m}{n}{
  <-> s*[1.1]  gsolomos8r
}{}

\newcommand{\vkappa}{\text{\usefont{U}{solomos}{m}{n}\symbol{'153}}}

   \usepackage{stmaryrd}

\usepackage{tikz}
\usetikzlibrary{arrows.meta}

\usepackage{indentfirst}

\usepackage{tocloft}
\let \savenumberline \numberline
\def \numberline#1{\savenumberline{#1.}}

\usepackage{etoolbox}
\patchcmd{\tableofcontents}{\@starttoc}{\vspace{-0.3cm}\@starttoc}{}{}

\usepackage{accents}
\newcommand\thickbar[1]{\accentset{\rule{.7em}{.4pt}}{#1}}
\usepackage{hyperref}
\hypersetup{
colorlinks=true,
linkcolor=violet,
citecolor=violet,
filecolor=purple,
urlcolor=cyan,
breaklinks=true
}

\begin{document}

\title{\vspace{-1.0cm} \textbf{Supersymmetric Grassmannian \\Sigma Models in Gross-Neveu Formalism}
\vspace{0.5cm}
}

\author{Dmitri Bykov$^{\,a,\,b,\,c\,}$\footnote{Emails:
 bykov@mi-ras.ru, dmitri.v.bykov@gmail.com} \qquad\qquad Viacheslav Krivorol$^{\,a,\,b,\,}$\footnote{Emails:
 vkrivorol@itmp.msu.ru, v.a.krivorol@gmail.com}
\\  \vspace{0cm}  \\
{\small $a)$ \emph{Steklov
Mathematical Institute of Russian Academy of Sciences,}} \\{\small \emph{Gubkina str. 8, 119991 Moscow, Russia} }\\
{\small $b)$ \emph{Institute for Theoretical and Mathematical Physics,}} \\{\small \emph{Lomonosov Moscow State University, 119991 Moscow, Russia}}\\
{\small $c)$ \emph{Moscow Institute of Physics and Technology,}} \\
{\small \emph{Institutskii per. 9, 141702  Dolgoprudny, Russia}}}

\date{}

{\let\newpage\relax\maketitle}

\maketitle

\vspace{0cm}
\textbf{Abstract.} 
We revisit the classical aspects of $\mathcal{N}=(2,2)$ supersymmetric sigma models with Hermitian symmetric target spaces, using the so-called Gross-Neveu (``first-order GLSM'') formalism. We reformulate these models for complex Grassmannians in terms of  simple supersymmetric Lagrangians with polynomial interactions. For maximal isotropic Grassmannians we propose two types of equivalent Lagrangians, which make either supersymmetry or the geometry of target space manifest. These reformulations can be seen as current-current deformations of curved $\beta \gamma$ systems. The~$\mathsf{CP}^{1}$ supersymmetric sigma model is our prototypical example.

\newpage
\tableofcontents

\newpage

\section{Introduction}
Grassmannian Sigma models (originally introduced in \cite{Macfarlane:1979hi, EICHENHERR1979}, see also \cite{CREMMER1978}) have long been a topic of interest, as they are known to have many similarities with four-dimensional gauge theories. For example, they exhibit asymptotic freedom, have instanton solutions \cite{DADDA1978,Atiyah:1984tk,Dunne:2015ywa}, chiral symmetry breaking, confinement \cite{DADDA1979}, etc. There is an extensive body of literature on sigma models, both classical and quantum, whose target spaces are Grassmann manifolds. Classical solutions of such models have been widely studied in the bosonic \cite{Macfarlane:1979hi,SASAKI1983,Delisle:2012qm,Din:1981bx,Sergeev} as well as in the supersymmetric case \cite{Sasaki1984,Hussin:2017fau} (see also~\cite{Bak:2006qk,Nakajima_2009}). In \cite{PhysRevResearch.1.023002} the authors explain how Grassmannian sigma models appear in the context of non-Abelian strings in four-dimensional supersymmetric gauge theories. The paper \cite{Kreshchuk:2019rom} focuses on the $(0, 2)$ deformation of the Grassmannian sigma model. Sigma models on Hopf-like fibrations over the Grassmannians were discussed in \cite{Schubring:2018pws}.  The phase structure of Grassmannian sigma models defined on a space  interval was studied in~\cite{Pavshinkin:2019bed}. Some applications in condensed matter physics can be found in~\cite{PhysRevB.99.165142}.   Finally, ``exotic'' Grassmannian sigma models\footnote{In the terminology of this work the term refers to orthogonal, symplectic and exceptional Grassmanians.} were considered in \cite{DonagiSharpe,Higashijima}.

In the present paper we continue investigating the so-called \textit{Gross-Neveu} \textit{formalism} (GN) with applications to supersymmetric Grassmannian sigma models. The GN approach resolves the issue of highly nonlinear nature of sigma models by introducing an equivalent Lagrangian with only polynomial interactions. Essentially, the Gross-Neveu approach combines the gauged linear sigma model (GLSM) construction with a first-order formulation within the context of sigma models. First-order formulations of sigma models can be traced back to~\cite{LosevFirstOrder} (see also \cite{Gamayun:2009sp,Zeitlin:2007mj,Gamayun:2009hy,Gamayun:2023atu}). 
The GLSM construction in the $\mathsf{CP}^{n-1}$ case  goes back to the work~\cite{CREMMER1978, DADDA1978, DADDA1979}; in the $\mathcal{N}=2$ supersymmetric setup the construction was widely used in~\cite{Salomonson, Witten:1993yc, MorrisonPlesser, NekrasovShatashvili} and many other works. 
Another, more recent, aspect of the story~\cite{BykovSigmaModelsAsGrossNeveuModels, bykov2023sigma2} is that the Gross-Neveu construction can be viewed as a deformation of curved $\beta\gamma$-systems~\cite{nekrasov2005lectures,Witten:2005px} by current-current interactions. 

The goal of this paper is to include supersymmetry in the Gross-Neveu framework. In particular, we will construct supersymmetric extensions of sigma models with symmetric Grassmannian target spaces within this formalism. Bosonic theory for Grassmannian sigma models in the Gross-Neveu formalism was developed in \cite{BykovKrivorolGSM},  the supersymmetric $\mathsf{CP}^{n-1}$ model was studied in \cite{BykovCpSUSY}, and various limits of the deformed supersymmetric $\mathsf{CP}^1$ model (including the most basic one, the super-Thirring model) were discussed in \cite{BykovPribytok}. 
This work is a logical continuation of those studies. By ``symmetric Grassmannians'' we refer to the following  Hermitian symmetric spaces:
\begin{itemize}
    \item all unitary Grassmannians $\mathsf{Gr}(m,n)={\mathbf{U}(n)\over \mathbf{U}(m)\times \mathbf{U}(n-m)}$,
    \item the extreme (\text{i.e.} maximal and minimal) orthogonal Grassmannians ${\mathsf{OGr}(1,n)={\mathbf{O}(n)\over \mathbf{O}(n-2)\times \mathbf{U}(1)}}$ and $\mathsf{OGr}(m,2m)={\mathbf{O}(2m)\over \mathbf{U}(m)}$.
    
    It is worth noting that $\mathsf{OGr}(1,n)$ is the non-singular quadric in $\CP^{n-1}$, i.e. the variety
    \begin{equation}\label{quadricdef}
        \sum_{i=1}^n X_i^2=0\,.
    \end{equation}
    Clearly, $\mathbf{O}(n)$ acts transitively on it, the stabilizer of a given point $(1, i, 0, \ldots , 0)$ being $\mathbf{O}(n-2)\times \mathbf{SO}(2)$.  In the definition of the quadric~(\ref{quadricdef}) we have picked the matrix of the quadratic form to be $\mathds{h}_{n}=\mathds{1}_{n}$; in what follows we will sometimes relax this assumption and assume that $\mathds{h}_{n}$ is an arbitrary symmetric non-degenerate matrix. 
    \item the maximal symplectic Grassmannian (or Lagrangian Grassmannian) $\mathsf{SGr}(m,2m):=\mathsf{LGr}(m,2m)={\mathbf{Sp}(2m)\over \mathbf{U}(m)}$.
\end{itemize}
It is well-known that these are the only Hermitian symmetric spaces that are homogeneous spaces of classical groups\footnote{There are two additional Hermitian symmetric spaces that are constructed using the exceptional groups $\mathbf{E}_6$ and $\mathbf{E}_7$.}. From the standpoint of the Gross-Neveu formalism these symmetric spaces are characterized by the fact that the  sigma model metrics that appear in second-order formalism after elimination of auxiliary fields are K\"ahler~\cite{BykovKrivorolGSM}. Thus, by the well-known result~\cite{Zumino:1979et}, such sigma models admit $\mathcal{N}=(2, 2)$ supersymmetric extensions. Conventional superfield formulations of sigma models with Grassmannian target spaces were studied in~\cite{Higashijima, SharpeExotic}. In the present paper we will explain how one can supersymmetrize the  GN Lagrangians  corresponding to symmetric Grassmannian sigma models. In the case of non-symmetric Grassmannians  we will describe obstructions to the supersymmetrization of the relevant GN systems: on the geometric side they are related to the fact that the target space metrics are not K\"ahler. 

The second part of this paper is dedicated to the geometric interpretation of the resulting supersymmetric Gross-Neveu Lagrangians. Recall that, in supersymmetric sigma models, fermions take values in the pullback of the tangent bundle to the target space~\cite{Hori:2003ic}. A fact from algebraic geometry is that, in the case of maximal orthogonal and symplectic Grassmannians, the tangent bundle is simply related to the tautological bundle -- it is either a symmetric or skew-symmetric square of its dual. 
It turns out that this leads to remarkable reformulations  for our Grassmannian Lagrangians. 
Surprisingly, one can introduce new fields in such a way that fermions would couple to the bosonic fields minimally, that is, only via the interaction with gauge fields. In particular, fermions are not included in the interaction term at all. 
However, the price one pays is that supersymmetry looks more complicated in this language. 
We explore the classical side of this issue and leave the study of the equivalence of the resulting formulations at the quantum level for future work.

The paper is organized as follows. In Sections \ref{GrSec} and \ref{OGrSec} we construct the supersymmetric Gross-Neveu formulations for unitary and extreme orthogonal/ Lagrangian Grassmannians, respectively. In Section \ref{MinimalCoupling} we construct a minimal fermion coupling reformulation for our basic example -- the $\mathsf{CP}^1$ model, which is followed by a generalization to the case of $\mathsf{OGr}(2, 2m)$ and $\mathsf{LGr}(2, 2m)$. In Appendix \ref{MinFermCP1} we check that our models with minimal fermion coupling coincide with the well-known supersymmetric $\mathsf{CP}^1$ sigma model in the geometric formulation.
In Appendix \ref{SUSYtransapp} we derive the supersymmetry transformations for the minimal fermion coupling models and check that the corresponding actions are invariant w.r.t. supersymmetry. Finally, in Appendix \ref{TungentBundles} we review how tangent bundles of Grassmanians may be expressed in terms of tautological bundles. The latter facts are widely used in the construction of supersymmetric models with minimal fermion coupling, and in fact serve, to a large extent, as  motivation for those constructions.

\section{Unitary Grassmannians $\mathsf{Gr}(m,n)$}\label{GrSec}
We start by recalling some basic geometric facts about Grassmannians. A point of the Grassmannian $\mathsf{Gr}(m,n)$ is an $m$-dimensional hyperplane $\mathcal{H}$ in $\mathbb{C}^n$. A given $m$-plane may be parametrized by a set of linearly independent  vectors $\{u_1,\ldots,u_m\}$ such that $\mathcal{H}=\mathrm{span}\{u_1,\ldots,u_m\}$. This parametrization is unique only up to taking linear combinations of these vectors, i.e. up to the natural action of $\mathbf{GL}(m,\mathbb{C})$. If we package the vectors in a single matrix\footnote{In the described setup the matrix $U$ obviously has maximal rank $m$. This will play a role in our discussion of supersymmetry later on.} as 
\begin{gather}
    U = (u_1,\ldots,u_m)
\end{gather} and identify $U\sim Ug$, where $g\in \mathbf{GL}(m,\mathbb{C})$, this equivalence class naturally defines an $m$-plane in $\mathbb{C}^n$. In this paper we will adopt precisely this gauge setup.  
In field theory language the identification above is a gauge transformation of the field~$U$. Gauge fixing corresponds to picking a section of the principal $\mathbf{GL}(m,\mathbb{C})$-bundle $\mathsf{St}(m,n)\xrightarrow[]{\pi}\mathsf{Gr}(m,n)$, where $\mathsf{St}(m,n)$ is the non-compact \textit{Stiefel manifold}\footnote{By definition, the non-compact Stiefel manifold $\mathsf{St}(m,n)$ is the set of all $m$-frames in $\mathbb{C}^n$. Packing the vectors of frames into matrices, one can  view the non-compact Stiefel manifold as the space of rank-$m$ matrices of size $m\times n$.}
and the natural projection $\pi$ maps an $m$-frame to the $m$-hyperplane spanned by this frame. Global sections do not exist, since the bundle is non-trivial, but one can pick a section locally.

Having this setup in mind, we first recall the bosonic sigma model with Grassmannian target space
\cite{BykovKrivorolGSM}
\begin{equation}
\label{ActionBosonicGR}
\mathcal{S}[U,V]=\int\mathrm{d}^2 z\, \left[2\Big(\mathrm{Tr}\big(V\thickbar{D}U\big)+\mathrm{Tr}\big(\thickbar{U}D\thickbar{V}\big)\Big)+\frac{\vkappa}{2\pi}\mathrm{Tr}\left(J_{\mathsf{sl}}\thickbar{J}_{\mathsf{sl}}\right)\right].
\end{equation}
Here $V$ and $U$ are $m\times n$ and $n\times m$ matrix fields respectively, $\thickbar{D} U=\thickbar{\partial}U-iU\thickbar{\mathcal{A}}$ and $\thickbar{D} V=\thickbar{\partial}U+i\thickbar{\mathcal{A}}\,V$ are the $\mathbf{GL}(m,\mathbb{C})$ covariant derivatives with the gauge field\footnote{We always assume that the action of the covariant derivative on the fields is consistent with the representations of these fields.} $A:=\mathcal{A}\mathrm{d}z+\thickbar{\mathcal{A}}\mathrm{d}\bar{z}$ and $J_{\mathsf{sl}}=2\pi\,UV$ is the global $\mathbf{GL}(n,\mathbb{C})$ current. For purposes of this work we assume that the worldsheet is flat. However, global effects on nontrivial worldsheets are of interest as well (see \cite{BykovRiemann}).

To construct the sypersymmetric version of this model we use the technique from~\cite{BykovCpSUSY}, where the $\mathsf{CP}^{n-1}$ supersymmetric sigma model was discussed in detail. In fact, all calculations from the $\mathsf{CP}^{n-1}$ case carry over to the case of unitary Grassmannians almost unchanged. Nevertheless, for the sake of self-consistency we will start by recalling the main ideas and carrying out similar calculations with minor~modifications. 

\subsection{The free system.}

It is convenient to start by considering the case of ``infinite volume $\vkappa=0$ and flat phase space (i.e. $A=0$)'' for the model (\ref{ActionBosonicGR}). To supersymmetrize the model, we consider the action
\begin{equation}
\label{FlatGrAction}
\mathcal{S}[U,V,B,C]=\int\mathrm{d}^2 z\, \Big[2\,\mathrm{Tr}\big(V\,\thickbar{\partial}U+\thickbar{U}\partial\thickbar{V}+B\,\thickbar{\partial}C-\thickbar{C}\partial\thickbar{B}\,\big)\Big]\,,
\end{equation}
where $V$ and $U$ are bosonic $m\times n$ and $n\times m$ matrix variables as before, whereas $B$ and $C$ are fermionic superpartners. 
This action is known to admit $\mathcal{N}=(2,2)$ supersymmetry \cite{Friedan:1985ge} (see also~\cite{Kapustin, Policastro})  
\begin{equation}
\label{SusyAction}
\delta U = \epsilon_1\, C,\quad \delta C = -\epsilon_2 \,\partial U\qquad\text{and}\qquad\delta V=\epsilon_2\, \partial B,\quad \delta B = -\epsilon_1\, V.
\end{equation}
The next step is to modify this free model in order to describe curved phase spaces. To this end we simply replace ordinary derivatives with covariant ones  
$\partial\rightarrow D$, $\thickbar{\partial}\rightarrow\thickbar{D}$ in the action (\ref{FlatGrAction}) and in the SUSY transformations (\ref{SusyAction}).
The e.o.m. for the gauge field $\thickbar{\mathcal{A}}$ is
\begin{eqnarray}\label{constr1}
    VU+BC = 0\,.
\end{eqnarray}
In mathematical terminology, this is a Hamiltonian constraint generating complex symplectic reduction. Although this interpretation is not strictly needed for the purposes of the present paper, it is useful to keep it in mind, cf.~\cite{BykovCpSUSY, BykovKrivorolGSM} for more details.

The new action is no longer supersymmetric though, since the SUSY variation is\footnote{We use the symbol $\simeq$ to indicate equivalence up to equations of motions and total derivatives.}
\begin{equation}
\delta\mathcal{S}\simeq \epsilon_2 \int \mathrm{d}^2z\,\big[\thickbar{D},D\big]\,BU\,,
\end{equation}
where $\big[\thickbar{D},D\big]$ is the commutator of covariant derivatives, i.e. the curvature of the gauge connection. The way to fix this supersymmetry obstruction is to impose a new constraint $BU= 0$ with the help of a new fermionic gauge field~$\thickbar{\mathcal{W}}$. Thus we come to consider the action 
\begin{equation}
\label{MiddleActionGr}
\mathcal{S}=\int\mathrm{d}^2 z\, \Big[2\,\mathrm{Tr}\big(V\thickbar{D}U+B\thickbar{D}C\big)+i\thickbar{\mathcal{W}}BU-c.c.\Big]\,.
\end{equation}
This theory is already supersymmetric. In particular, the system of constraints 
is closed w.r.t. SUSY transformations, i.e.
\begin{equation}
\delta(BU)=-\epsilon_1(VU+BC)\simeq 0 \quad\text{and}\quad\delta(VU+BC)=\epsilon_2\,D(BU)\simeq 0. 
\end{equation}

\subsection{Interactions.} 
Now we supersymmetrize the interaction term
\begin{equation}
\label{NonSUSYinteractionGr}
\mathcal{S}_{\text{int}}=\frac{\vkappa}{2\pi} \int\mathrm{d}^2 z\, \mathrm{Tr}\left(J_{\mathsf{sl}}\thickbar{J}_{\mathsf{sl}}\right)=2\pi\vkappa\int\mathrm{d}^2 z\, \mathrm{Tr}\big(UV\thickbar{V}\thickbar{U}\big)\,.
\end{equation}
The global $\mathfrak{sl}_n$ current of the free supersymmetric system~(\ref{FlatGrAction}) is
\begin{equation}
    \mathcal{J}_{\mathsf{sl}}=2\pi\left(UV-CB\right)\,, 
\end{equation}
so that we need to modify the interaction (\ref{NonSUSYinteractionGr}) by $J_{\mathsf{sl}}\rightarrow\mathcal{J}_{\mathsf{sl}}$. 

Thus, we assume that the final form of the  supersymmetric action is
\begin{align}
\label{FinalActionGr}
\mathcal{S}=\int\mathrm{d}^2 z\, \Big[2\,\mathrm{Tr}\big(V\thickbar{D}U+B\thickbar{D}C+i\thickbar{\mathcal{W}}BU\big)-c.c.\Big]+\frac{\vkappa}{2\pi} \int\mathrm{d}^2 z\, \mathrm{Tr}\left(\mathcal{J}_{\mathsf{sl}}\thickbar{\mathcal{J}}_{\mathsf{sl}}\right)\,.
\end{align}
Let us prove that this action is really supersymmetric. The SUSY variation of the current is
\begin{equation}
\label{VariationJGr}
\delta\mathcal{J}_{\mathsf{sl}}=\epsilon_2\partial\widetilde{\mathcal{J}_{\mathsf{sl}}},\quad\text{where}\quad\widetilde{\mathcal{J}_{\mathsf{sl}}}=2\pi\, UB\,.
\end{equation}
Also notice that the e.o.m. imply that
\begin{equation}
\label{PCMlikeEOM}
\thickbar{\partial}\mathcal{J}_{\mathsf{sl}}=\frac{\vkappa}{2\pi}\big[\mathcal{J}_{\mathsf{sl}},\thickbar{\mathcal{J}}_{\mathsf{sl}}\big]\,. 
\end{equation}
Using (\ref{VariationJGr}) and (\ref{PCMlikeEOM}), it is easy to see that the interaction term is supersymmetric. Its variation can be calculated as follows:
\begin{equation}
\label{VariationOfInteraction}
\delta\mathrm{Tr}\left(\mathcal{J}_{\mathsf{sl}}\thickbar{\mathcal{J}}_{\mathsf{sl}}\right)=\epsilon_2\mathrm{Tr}\left(\thickbar{\mathcal{J}}_{\mathsf{sl}}\,\delta\widetilde{\mathcal{J}_{\mathsf{sl}}}\right)+c.c.\simeq \epsilon_2\,\frac{\vkappa}{2\pi}\mathrm{Tr}\Big(\big[\mathcal{J}_{\mathsf{sl}},\widetilde{\mathcal{J}_{\mathsf{sl}}}\big]\,\thickbar{\mathcal{J}}_{\mathsf{sl}}\Big)+c.c.\,.
\end{equation}
The calculation of the commutator gives
\begin{equation}
\label{CommutatorSl}
\big[\mathcal{J}_{\mathsf{sl}},\widetilde{\mathcal{J}_{\mathsf{sl}}}\big]=4\pi^2\Big(U(VU+BC)B-UBUV-CBUB\Big)\simeq 0\,,
\end{equation}
where we have taken advantage of the constraints imposed by the gauge fields $\thickbar{\mathcal{A}}$, $\thickbar{\mathcal{W}}$. 
The same applies to complex conjugate terms. It means that the SUSY variation of the action $\delta\mathcal{S}\simeq 0$, thus the action (\ref{FinalActionGr}) is on-shell supersymmetry invariant under the transformations 
\begin{align}
\label{SusyActionFinal}
\delta U = \epsilon_1\, C,\quad \delta C = -\epsilon_2 \,D U\qquad\text{and}\qquad\delta V=\epsilon_2\, D B,\quad \delta B = -\epsilon_1\, V.
\end{align}

\subsection{Gauge symmetry.} \label{gaugesymmsec}
Let us briefly discuss the gauge symmetry of this model. The constraints give rise to the following gauge transformations (together with the standard compensating transformations of the gauge fields $\thickbar{\mathcal{A}}$ and $\thickbar{\mathcal{W}}$):
\begin{align}
&U\rightarrow Ug\,,\quad V\rightarrow g^{-1}V+e\,B\,,\\
&C\rightarrow Cg+Ue\,,\quad B\rightarrow g^{-1}B\,,\\
&\textrm{where}\quad g\in \mathbf{GL}(m,\mathbb{C})\,,\quad e\in \mathsf{\Pi}\mathrm{Mat}_{m}\nonumber\,.
\end{align}
Here $\mathrm{Mat}_{m}$ is the additive group of $m\times m$ matrices, whereas the symbol `$\mathsf{\Pi}$' means that the corresponding parameters are Grassmann numbers (i.e., it indicates the shift of  fermion grading). 
Thus, the gauge group (or group of complex symplectic reduction)~$\mathcal{G}$ is 
\begin{equation}\label{gaugegroup}
    \mathcal{G}=G\ltimes\mathsf{\Pi} \mathfrak{g}\,,
\end{equation}
where in this case $G=\mathbf{GL}(m,\mathbb{C})$, and $\mathfrak{g}=\mathrm{Mat}_m$ is its Lie algebra.

\vspace{0.3cm}
The structure~(\ref{gaugegroup}) of the gauge group is rather general, and, in particular, also holds for orthogonal and symplectic Grassmannians, as we shall see below. Let us explain why this is always the case for models of this type. To this end, we need a few facts about the superspace formulation of these models. Although the superspace form of interacting models has not yet been developed, the holomorphic part of the free system~(\ref{FlatGrAction}) (which is all we need for describing the gauge fields) may be easily written in $\mathcal{N}=(2, 2)$ superspace~\cite{BykovPribytok}. Here the superspace coordinates are $\theta_1, \theta_2, \thickbar{\theta}_1, \thickbar{\theta}_2$, and the Lagrangian reads
\begin{eqnarray}
 \mathcal{L}_{\mathrm{hol}}= \int\,d^3\theta \,\mathbf{U} \mathbf{B}\,,
\end{eqnarray}
where $\mathbf{U}$ is a chiral superfield (depending non-trivially on $\theta_1, \thickbar{\theta}_2$) and $\mathbf{B}$ is twisted chiral (depending non-trivially on $\theta_1, \thickbar{\theta}_1$). The integration is over the superspace coordinates on which the integrand depends non-trivially, i.e. $d^3\theta= d\theta_1 d\thickbar{\theta}_1 d\thickbar{\theta}_2$. Suppose the above system is invariant w.r.t. some global (complex) group $G$, for example 
$\mathbf{B}\to g \mathbf{B}$, $\mathbf{U} \to \mathbf{U} g^{-1}$, where $g\in G$. To gauge it, we should allow $g$ to depend on $z, \bar{z}$. However, to preserve chirality properties of the fields one should require that
\begin{eqnarray}\label{BUgaugetrans}
    \mathbf{B}\to g_{\mathrm{tc}}(z, \bar{z})\mathbf{B}\,, \quad \mathbf{U} \to \mathbf{U} g_{\mathrm{c}}^{-1}(z, \bar{z})\,,
\end{eqnarray}
 with $g_{\mathrm{c}}$ and $g_{\mathrm{tc}}$ respectively chiral and twisted chiral fields. To mantain invariance of the Lagrangian, one also introduces a \emph{semi-chiral} group-valued gauge superfield $\mathbf{g_V}$; in standard physics conventions it might perhaps be more common to write it as $\mathbf{g_V}=e^{\mathbf{V}}$, with $\mathbf{V}$ a Lie algebra-valued gauge superfield. Either way, the gauge-invariant Lagrangian has the form
\begin{eqnarray}
 \mathcal{L}_{\mathrm{gauged}}= \int\,d^3\theta \,\mathbf{U} \,\mathbf{g_V}\, \mathbf{B}\,.
\end{eqnarray}
On top of~(\ref{BUgaugetrans}) one should postulate the transformation law of the gauge field:
\begin{equation}\label{twchiralgaugetrans}
    \mathbf{g_V}\to g_{\mathrm{c}}(z, \bar{z})\,\mathbf{g_V}\,g_{\mathrm{tc}}^{-1}(z, \bar{z})\,.
\end{equation}
Just like in more conventional supersymmetric models, here one can use part of this gauge symmetry to simplify the form of $\mathbf{g_V}$ by bringing it to generalized Wess-Zumino (WZ) gauge. By a slight generalization of the argument in~\cite{BykovPribytok}, the gauge field in WZ gauge has the form (here $z_-=z-{1\over 2} \theta_1 \theta_2$):
\begin{eqnarray}\label{WZgauge}
\mathbf{g_V}=\mathds{1}+\thickbar{\theta}_1 \thickbar{\theta}_2 \thickbar{\mathcal{A}}(z_-, \bar{z})+\theta_1 \thickbar{\theta}_1 \thickbar{\theta}_2 \thickbar{\mathcal{W}}(z_-, \bar{z})\,,
\end{eqnarray}
where, as we shall momentarily see, $\thickbar{\mathcal{A}}$ and $\thickbar{\mathcal{W}}$ are the gauge fields for the two factors in~(\ref{gaugegroup}), each of them taking values in $\mathfrak{g}$.

The question is what gauge transformations remain in the WZ gauge~(\ref{WZgauge}), i.e. for which choice of $g_{\mathrm{t}}$ and $g_{\mathrm{tc}}$ in~(\ref{twchiralgaugetrans}) the transformed field remains in the form~(\ref{WZgauge}). A direct inspection of the $\theta$-dependence of various superfields involved shows that $g_{\mathrm{c}}$ and $g_{\mathrm{tc}}$ should be of the form
\begin{eqnarray}
g_{\mathrm{c}}=g(z_-, \bar{z}_+)\big(\mathds{1}+\theta_1\,\chi(z_-, \bar{z}_+)\big)\,,\\
g_{\mathrm{tc}}=g(z_-, \bar{z}_-)\big(\mathds{1}+\theta_1\,\chi(z_-, \bar{z}_-)\big)\,.
\end{eqnarray}
It is already clear from this form that the gauge group is~(\ref{gaugegroup}) since  two consecutive gauge transformations give
\begin{eqnarray}
(g_{\mathrm{c}})_1 (g_{\mathrm{c}})_2=g_1\,(\mathds{1}+\theta_1 \chi_1)\,g_2\,(\mathds{1}+\theta_1 \chi_2)=g_1 g_2 \, \Big(\mathds{1}+\theta_1\left(g_2^{-1}\chi_1 g_2+ \chi_2\right)\Big)\,,
\end{eqnarray}
which is the group composition law of the semi-direct product. One could as well compute the action of the residual gauge transformations on the component gauge fields $\thickbar{\mathcal{A}}, \thickbar{\mathcal{W}}$ using~(\ref{twchiralgaugetrans}). This gives
\begin{eqnarray}
&&\thickbar{\mathcal{A}} \to g\thickbar{\mathcal{A}}g^{-1}+\thickbar{\dd} g\,g^{-1}\,,\\
&&\thickbar{\mathcal{W}} \to g\Big(\thickbar{\mathcal{W}}+\thickbar{\dd}\chi+\big[\,\chi, \thickbar{\mathcal{A}}\,\,\big] \Big) g^{-1}\,,
\end{eqnarray}
which again reflects the semi-direct product structure.

\vspace{0.3cm}
In this section, we have proven \emph{on-shell} invariance w.r.t. supersymmetry transformations, i.e., in the proof we have made use of the e.o.m. for the currents~(\ref{PCMlikeEOM}) as well as the gauge constraints. It is also possible to find off-shell transformations by suitably adapting the method described in Appendix A of \cite{BykovCpSUSY}. 
We leave details of the derivation to the interested reader.

\section{Extreme Grassmannians}\label{OGrSec}
In this section we discuss the supersymmetrization of  sigma models with target spaces $\mathsf{OGr}(m,n)$ and $\mathsf{SGr}(m,n)$. The main claim is that here, in contrast to the unitary case, there is a non-trivial obstruction to supersymmetry that disappears only in the extreme cases $\mathsf{OGr}(1,n)$, $\mathsf{OGr}(m,2m)$ and $\mathsf{LGr}(m,2m)$. This is in line with well-known facts about supersymmetry in the conventional geometric (second-order) formulation. As is well-known, it is sigma models with Kähler target spaces that allow for extended $\mathcal{N}=(2, 2)$ supersymmetry~\cite{Zumino:1979et}. However, the GN formalism produces Kähler metrics on Grassmannians only if these are symmetric spaces\footnote{In other cases the \emph{normal} metric that arises out of the GN formulation is not K\"ahler.}, which is in general not the case for orthogonal and symplectic Grassmannians. 

\subsection{Orthogonal Grassmannians.} We begin by considering the orthogonal case and then switch to the symplectic case later on, since the two are related by small modifications. The technology for the supersymmetrization of the actions is to a large extent similar to the one used above for unitary Grassmannians. Therefore, we will mainly focus on the novel aspects that arise. 

The action of the purely bosonic orthogonal Grassmannian $\mathsf{OGr}(m,n)$ sigma models in Gross-Neveu formalism is\footnote{In \cite{BykovKrivorolGSM} the gauge field $\thickbar{\mathcal{R}}$ is denoted by $\thickbar{\mathcal{A}}_+$.} \cite{BykovKrivorolGSM}
\begin{equation}
\label{ActionBosonicOGR}
\mathcal{S}[U,V]=\int\mathrm{d}^2 z\, \mathrm{Tr}\left[\Big(2V \circ\thickbar{D}U+i\thickbar{\mathcal{R}}\,U^t\circ U-c.c.\Big)+\frac{\vkappa}{4\pi}\mathrm{Tr}\left(J_{\mathsf{so}}\thickbar{J_{\mathsf{so}}}\right)\right].
\end{equation}
Here by $\circ$ we denote the scalar product with insertion of a 
symmetric non-degenerate matrix $\mathds{h}_n$ of size $n\times n$ satisfying\footnote{Here and hereafter $\mathds{1}_n$ is the unit matrix of size $n\times n$.} $\mathds{h}_n^2=\mathds{1}_n$. For example, $V\circ U=V\mathds{h}_n U$. The matrix $\mathds{h}_n$ represents, in a given basis, a symmetric bilinear form in $\mathbb{C}^n$. The simplest choice is $\mathds{h}_n = \mathds{1}_n$, and we often use it, but sometimes we will keep the matrix arbitrary to demonstrate the parallels between the orthogonal and symplectic cases (in a nutshell, in order to switch to the symplectic case one simply replaces $\mathds{h}_n$ by a skew-symmetric matrix). Besides,   
\begin{gather}
J_{\mathsf{so}}=2\pi\big(UV-V^tU^t\big)\mathds{h}_n
\end{gather}
is the $\mathbf{SO}(n,\mathbb{C})$ global current.  Notice also that the important difference of the above action from~(\ref{ActionBosonicGR}) is in the presence of an extra gauge field $\thickbar{\mathcal{R}}$ imposing the isotropy constraint.

As in the previous section, we start from the free unconstrained system (\ref{FlatGrAction}),  with
the supersymmetry transformations given by~(\ref{SusyAction}). Next, we impose the isotropy constraints characteristic of the  orthogonal Grassmannian. As in the unitary case, supersymmetrization of the $\thickbar{\mathcal{A}}$-constraint $V\circ U+B\circ C = 0$ requires introducing a superpartner gauge field $\thickbar{\mathcal{W}}$ imposing the superpartner constraint $B\circ U = 0$. Since the supersymmetry variation of the isotropy constraint $U^t\circ U = 0$ imposed by the bosonic gauge field $\thickbar{\mathcal{R}}$ is
\begin{equation}
\label{variationUtU}
\delta(U^t\circ U)=\epsilon_1\left(C^t\circ U+U^t\circ C\right)\,,
\end{equation}
one should add the superpartner constraint $C^t\circ U+U^t\circ C=0$ to the action. The corresponding Lagrange multiplier is denoted by $\thickbar{\mathcal{Q}}$ and plays the role of a fermionic gauge field.

In the last step we need to add the interaction. In the orthogonal case we modify the current-current interaction of the bosonic model~(\ref{ActionBosonicOGR}) by the replacement $J_{\mathsf{so}}\rightarrow\mathcal{J}_{\mathsf{so}}$, where
\begin{equation}
\label{CurrentSO}
\frac{1}{2\pi}\mathcal{J}_{\mathsf{so}}=\big(UV-CB\big)\mathds{h}_n-\big(UV-CB\big)^t\,\mathds{h}_n\,.
\end{equation}
The e.o.m. for $\mathcal{J}_{\mathsf{so}}$  have the same form (\ref{PCMlikeEOM}), as before.
The SUSY variation of this current is
\begin{equation}
\label{VariationJOGr}
\delta\mathcal{J}_{\mathsf{so}}=\epsilon_2\partial\widetilde{\mathcal{J}_{\mathsf{so}}},\quad\text{where}\quad\widetilde{\mathcal{J}_{\mathsf{so}}}=2\pi\big(UB-B^tU^t\big)\mathds{h}_n\,.
\end{equation}
Like in the unitary case \big(see (\ref{VariationOfInteraction})\big), the variation of the interaction term is proportional to the commutator $\big[\mathcal{J}_{\mathsf{so}},\widetilde{\mathcal{J}_{\mathsf{so}}}\big]$. In the unitary case this commutator is proportional to  the equations of motion, but for orthogonal Grassmannians it is not true in general. 
An explicit evaluation of the commutator gives the following result: 
\begin{equation}
\label{CurrentsCommutatorOGr}
\big[\mathcal{J}_{\mathsf{so}},\widetilde{\mathcal{J}_{\mathsf{so}}}\big] \simeq 4\pi^2\Big(CB\circ B^tU^t+UB\circ B^tC^t+U\big(B\circ V^t-V\circ B^t\big)U^t\Big)\mathds{h}_n\,,
\end{equation} 
where we have used the constraints to simplify the r.h.s. 
Our statement is that $\big[\mathcal{J}_{\mathsf{so}},\widetilde{\mathcal{J}_{\mathsf{so}}}\big]\simeq 0$ for the   `minimal' and `maximal' orthogonal Grassmannians $\mathsf{OGr}(1,n)$ and $\mathsf{OGr}(m,2m)$. Let us prove this proposition.

\subsubsection{Quadric.} The case of the minimal Grassmannian is elementary, since   
here $B$ and $V$ are just fermionic and bosonic vectors, respectively. Due to the fact that $B_i$'s are Grassmann numbers, $B\circ B^t=0$. Besides, $B\circ V^t = V\circ B^t$, so that $\big[\mathcal{J}_{\mathsf{so}},\widetilde{\mathcal{J}_{\mathsf{so}}}\big]\simeq 0$.

\subsubsection{Maximal Grassmannian.} The case of the maximal Grassmannian is somewhat more involved. For convenience let us fix $\mathds{h}_n=\mathds{1}_n$ for the moment.  Recall that the columns of the matrix $U$   comprise $m$ isotropic vectors, $U^t U=0$. As usual, we assume that $U$ is of \emph{maximal rank}, i.e. that these vectors are linearly independent. The $m$ complex conjugate vectors, being columns\footnote{Here and hereafter, by the star $^\ast\,$ we denote ordinary complex conjugation (without transposition).} of $U^\ast$, complete the original $m$ vectors to a full basis. Indeed, these vectors span an $m$-plane orthogonal to the original one, since $\thickbar{U}^\ast U=U^t U=0$. The following partition of unity is a consequence of this fact:
\begin{equation}\label{partunity}
U\big(\thickbar{U}U\big)^{-1}\thickbar{U} + \thickbar{U}^t\big(\thickbar{U}U\big)^{-1t}U^t=\mathds{1}_{2m}\,.
\end{equation}
This statement is precisely that the sum of projectors to orthogonal subspaces is the identity operator. It is convenient to introduce the notation 
\begin{eqnarray}\label{Pdef}
P:=\big(\thickbar{U}U\big)^{-1}\thickbar{U}\,.
\end{eqnarray}
In these terms the partition of unity takes the form 
\begin{equation}
    UP+(UP)^t = \mathds{1}_{2m}\,.
\end{equation}

Now, substituting the partition of unity between the two $B$'s in $B B^t$, one finds that each of the two terms contains $B U$ and thus vanishes. This then automatically leads to $B V^t-V B^t\simeq 0$, because the two constraints are related by supersymmetry, \text{i.e.}
$\delta\big(B B^t\big) = \epsilon_1\big(B V^t-V B^t\big)$.
However this can also be proven directly by inserting the partition of unity twice and using the $\thickbar{\mathcal{W}}$-constraint $B U = 0$:
\bear
&&B V^t-V B^t=BP^t (VU)^t-VU PB^t\simeq\{\textrm{using}\; VU+BC\simeq 0\}\simeq \nonumber\\
&&\simeq BP^tC^tB^t+BCPB^t=BP^t\left(C^tU+U^tC\right)PB^t\simeq 0\,,
\eear
where in the last step we have used the $\thickbar{\mathcal{Q}}$-constraint $C^tU+U^tC = 0$ which follows from (\ref{variationUtU}). Thus the proof is complete.

Summarizing the above, we have constructed the supersymmetric formulation for sigma models with $\mathsf{OGr}(1,n)$ and $\mathsf{OGr}(m,2m)$ target spaces in Gross-Neveu formulation. The final action, invariant under SUSY transformations (\ref{SusyActionFinal}), in both cases can be written in the form
\begin{align}
\mathcal{S}=&\int\mathrm{d}^2 z\, \mathrm{Tr}\,\Big[2\big(V\circ \thickbar{D}U+B\circ \thickbar{D}C\,\big)+i\thickbar{\mathcal{R}}\,U^t\circ U+i\thickbar{\mathcal{W}}\,B\circ U+\nonumber\\ \label{FinalActionOGr}+&i\thickbar{\mathcal{Q}}\left(C^t\circ U+U^t\circ C\right)-c.c.\Big]+\frac{\vkappa}{4\pi} \int\mathrm{d}^2 z\, \mathrm{Tr}\left(\mathcal{J}_{\mathsf{so}}\thickbar{\mathcal{J}}_{\mathsf{so}}\right)\,,
\end{align}
where we have restored the matrix $\mathds{h}_n$ entering via the scalar product $\circ$. 
In the case of the quadric the Lagrangian can be simplified by getting rid of the trace and using the fact that the two terms in the $\thickbar{\mathcal{Q}}$-constraint are identical.

For orthogonal Grassmannian models, the  additional constraints (as compared to the unitary case) generate additional gauge transformations. As a result, 
the full gauge group acts on the fields as follows:
\begin{align}
&U\rightarrow Ug\,,\quad V\rightarrow g^{-1}V+q\,U^t+e\,B+y\,C^t\,,\label{GaugeGroupOGr1}\\
&C\rightarrow Cg+Ue\,,\quad B\rightarrow g^{-1}B+y\,U^t\,,\label{GaugeGroupOGr2}\\
&g\in \mathbf{GL}(m,\mathbb{C})\,,\quad q\in\mathrm{Mat}^{\text{symm}}_m\,,\quad e\in \mathsf{\Pi}\mathrm{Mat}_m\,,\quad y\in \mathsf{\Pi}\mathrm{Mat}^{\text{symm}}_m\,.\label{GaugeGroupOGr3}
\end{align}
Here $(g, e)$ and $(q, y)$ are superdoublets, corresponding to the gauge group structure~(\ref{gaugegroup}). From the standpoint of~(\ref{gaugegroup}), the bosonic group  is 
\begin{gather}\label{bosgaugegrouportho}
    G=\mathbf{GL}(m,\mathbb{C}) \ltimes \mathrm{Mat}^{\text{symm}}_m
\end{gather}
in this case.

\subsection{Non-maximal Grassmannians: obstructions to supersymmetry.} Let us note that in general (\text{i.e.} for arbitrary orthogonal Grassmannian) $BB^t$ and $BV^t-VB^t$ cannot be reduced to zero by the e.o.m. But one can ask, what if we add these constraints by hand, as we do for $BU$ and $C^tU+U^tC$? This can be done, but it ruins the geometric interpretation of the field $U$ as a set of coordinates on a Grassmannian. The reason is that the new constraints generate extra gauge transformations $U\rightarrow U+B^tk$, where $k$ is a fermionic $m\times m$ matrix. As a result, the phase space is no longer a cotangent bundle. 
More than that, if we relax the condition that $U$ has maximal rank, the above proof that $BB^t$ and $BV^t-VB^t$ vanish on-shell for the minimal and maximal orthogonal Grassmannians would fail as well, since in the proof we have used the partition of unity (\ref{partunity}). 

Thus, by weakening the condition on the rank of $U$ and/or adding additional constraints, we obtain self-consistent supersymmetric models of  Gross-Neveu type, which, however, apparently do not have a geometric interpretation.

\subsection{Gauge fixing.} 
The final point we wish to discuss here is the introduction of a convenient gauge. Following the general logic, we know that to get to the geometric formulation we need to integrate out the auxiliary fields (the `momenta' $V$ and $\thickbar{V}$, for example). But, as we discussed in \cite{BykovKrivorolGSM}, the $(V, \thickbar{V})$ quadratic form in the interaction part of  the action (\ref{FinalActionOGr}) is degenerate. This is true for the supersymmetric system as well. Therefore, similar to QCD, in order to integrate out $(V, \thickbar{V})$ we need to use the $\alpha$-gauge trick. The simplest suitable gauge for the bosonic\footnote{The fermionic superpartners for these constraints are $\thickbar{C}U = 0$ and $B^\ast U + (B^\ast U)^t = 0$.} sector is to fix 
\begin{gather}
    \thickbar{U}U=\mathds{1}_m\quad \quad \textrm{and}\quad\quad \mathcal{F}:=V^\ast U+(V^\ast U)^t = 0\,.
\end{gather}
The latter condition can be picked, since the gauge transformation $V\rightarrow V+q\,U^t$ shifts the symmetric part of $V^\ast U$ by a symmetric matrix~$q$. Although, as $\delta \mathcal{F} \neq 0$, this naively breaks supersymmetry, it can be restored by an \textit{additional gauge transformation}. 
Indeed, suppose the SUSY variation has the form $\delta\mathcal{F}=2R$, with $R$ a symmetric matrix. Then one can make an additional gauge transformation $\delta' V=-R^\ast U^t$, so that the modified SUSY transformation $\delta+\delta'$ would leave the gauge constraint unaltered. Of course, in our supersymmetric model (\ref{FinalActionOGr}) more gauge conditions need to be chosen. In addition to $\thickbar{U} U = \mathds{1}_{m}$ and $\mathcal{F} = 0$, one needs to introduce their corresponding superpartners. However, a similar argument can be made in those cases as well.

\subsection{Symplectic Grassmannians.}  
The supersymmetrization of symplectic Grassmannians $\mathsf{SGr}(m,2n)$ goes parallel to the orthogonal ones. In this section we elaborate some minor differences and provide the final answers. 
The action of purely bosonic symplectic Grassmannian $\mathsf{SGr}(m,2n)$ sigma models in Gross-Neveu formalism is \cite{BykovKrivorolGSM}
\begin{equation}
\label{ActionBosonicOGR}
\mathcal{S}[U,V]=\int\mathrm{d}^2 z\, \left[2\Big(\mathrm{Tr}\big(V\circ\thickbar{D}U\big)+i\thickbar{\mathcal{R}}\,U^t\circ U-c.c.\Big)+\frac{\vkappa}{4\pi}\mathrm{Tr}\left(J_{\mathsf{sp}}\thickbar{J}_{\mathsf{sp}}\right)\right].
\end{equation}
Throughout this section the circle $\circ$ stands for matrix multiplication with insertion of a $2n\times2n$ skew-symmetric non-degenerate matrix $\omega_{2n}$, which can be viewed as a symplectic form in $\mathbb{C}^{2n}$. The global current here is
\begin{equation}
    J_{\mathsf{sp}} = 2\pi\left(UV+V^tU^t\right)\omega_{2n}\,.
\end{equation}
Note that this action differs from (\ref{ActionBosonicOGR}) by the replacement\footnote{So that, for example, in this section $V\circ U$ means $V\omega_{2n}U$.} $\mathds{h}_n\rightarrow\omega_{2n}$ and an extra sign in the current. We are thus led to consider the following fermionic extension of the action \big(the symplectic counterpart of (\ref{FinalActionOGr})\big):
\begin{eqnarray}
&&\mathcal{S}=\int\mathrm{d}^2 z\, \mathrm{Tr}\,\Big[2\big(V\circ\thickbar{D}U+B\circ\thickbar{D}C\,\big)+i\thickbar{\mathcal{R}}\,U^t\circ U+i\thickbar{\mathcal{W}}\,B\circ U+\nonumber\\ \label{FinalActionSGr}
&&\quad\quad +i\thickbar{\mathcal{Q}}\left(C^t\circ U+U^t\circ C\right)-c.c.\Big]+\frac{\vkappa}{4\pi} \int\mathrm{d}^2 z\, \mathrm{Tr}\left(\mathcal{J}_{\mathsf{sp}}\thickbar{\mathcal{J}}_{\mathsf{sp}}\right).
\end{eqnarray}
Here the `supersymmetric' current is
\begin{gather}
    \mathcal{J}_{\mathsf{sp}}=2\pi\big(UV-CB\big)\omega_{2n}+2\pi\big(UV-CB\big)^t\omega_{2n}\,.
\end{gather}
As in the orthogonal case, the obstructions to  supersymmetric invariance of the action (\ref{FinalActionSGr}) are proportional to $B\circ B^t$ and its superpartner $B\circ V^t-V\circ B^t$.
The key difference between the symplectic and orthogonal cases is that the action for the minimal symplectic Grassmannian $\mathsf{SGr}(1,2n)$ is no longer supersymmetric\footnote{Except the case $n=1$, because $\mathsf{LGr}(1,2)\simeq\mathsf{CP}^1$. We discuss the $\mathsf{CP}^1$ model in detail is section~\ref{CP1}.}. The reason is that, although $B^t$ is a (fermionic) vector, the bilinear combination $B\circ B^t = B\omega_{2n}B^t$ in general no longer vanishes, since $\omega_{2n}$ is skew-symmetric. This was to be  expected, as $\mathsf{SGr}(1,2n)$ is not a symmetric space. 
As a result, the Gross-Neveu formalism does not produce a K\"ahler metric on the target space (see~\cite{BykovKrivorolGSM}) that is needed for supersymmetry. 
One can prove that $B\circ B^t$ and $B\circ V^t-V\circ B^t$ vanish (again, under the assumption that $U$ is of maximal rank) only in the maximal (Lagrangian)   case $\mathsf{SGr}(m,2m):=\mathsf{LGr}(m,2m)$\footnote{This is an expected result as well, since among symplectic Grassmannians only Lagrangian ones are Hermitian symmetric.}. 
The proof is parallel to the one for orthogonal Grassmannians from the previous section.

The gauge group of the model acts on the fields similarly to (\ref{GaugeGroupOGr1}-\ref{GaugeGroupOGr3}), the only difference being that in the symplectic case $$q\in\mathrm{Mat}^{\text{skew-symm}}_m\quad\text{and}\quad y\in \mathsf{\Pi}\mathrm{Mat}^{\text{skew-symm}}_m\,,$$
so that the corresponding matrices are skew-symmetric rather than symmetric.

\section{Maximal Grassmannians \textit{via} minimal fermion coupling}\label{MinimalCoupling}
In this section we focus on the geometric aspects of maximal Grassmannian sigma models. We show that going to geometry is not just an exercise in abstract nonsense, but  leads to very concrete field-theoretic results.

The standard formulation of supersymmetric sigma models (see Chapter 13 in \cite{Hori:2003ic}) with K\"ahler target spaces dictates that fermions must be sections of the tangent bundle of the target space (with shifted fermionic parity in each fiber). The naive explanation of this fact is that fermions carry vector indices, which contract with Christoffel symbols, Riemann tensor and so on.  

In terms of our formalism this fact has interesting consequences in the cases of maximal Grassmannians $\mathsf{OGr}(m,2m)$ and $\mathsf{LGr}(m,2m)$. A standard result from projective algebraic geometry states that there are isomorphisms of vector bundles (for details see Appendix \ref{TungentBundles})
\begin{equation}\label{tanbun}
\mathrm{T}\mathsf{OGr}(m,2m)\simeq\left(\bigwedge^2\mathcal{S}^\vee\right)\bigg|_{\mathsf{OGr}(m,2m)}\,,\quad 
\mathrm{T}\mathsf{LGr}(m,2m)\simeq \Big(\mathrm{Sym}^2\mathcal{S}^\vee\Big)\Big|_{\mathsf{LGr}(m,2m)}\,.
\end{equation}
Here $\mathrm{T}$ denotes the tangent bundle, restriction means pull-back bundle, $\mathcal{S}$ is the tautological bundle over the Grassmannian and $\mathcal{S}^\vee$ is the dual one (both of rank  $m$).

Since fermions take values in the tangent bundle, it follows from~(\ref{tanbun}) that 
they should be represented by skew-symmetric \big(in the  $\mathsf{OGr}(m,2m)$ case\big) and symmetric \big(in the  $\mathsf{LGr}(m,2m)$ case\big) $m\times m$ matrices. This nicely matches the counting of degrees of freedom in both theories. Taking into account the action of the gauge group (\ref{GaugeGroupOGr1}-\ref{GaugeGroupOGr3}) and its symplectic counterpart one easily finds the number of fermionic degrees of freedom in the orthogonal case to be $\frac{m(m-1)}{2}$ \big(and $\frac{m(m+1)}{2}$ in the symplectic case\big). Clearly, this coincides with the number of components of skew-symmetric and symmetric $m\times m$ matrices, respectively. 
However, in the Lagrangians (\ref{FinalActionOGr}) and (\ref{FinalActionSGr}) fermions\footnote{More precisely, in this introduction we discuss ``configuration space'' fermions. The full phase space, where all fields live, should be seen as the cotangent bundle  $\mathrm{T}^\ast\big(\mathsf{\Pi}\mathrm{T}\mathsf{OGr}(m,2m)\big)$ or $\mathrm{T}^\ast\big(\mathsf{\Pi}\mathrm{T}\mathsf{LGr}(m,2m)\big)$.} are $m\times 2m$ matrices, and the `tangent bundle structure' described above is hidden.
The main goal of this section is to construct a reformulation that reflects the tangent bundle geometry in a more explicit way.

Before passing to the general case, 
let us consider 
the simplest example of Lagrangian Grassmannian   $\mathsf{LGr}(1,2)\simeq \mathsf{CP}^1$ without interactions, \text{i.e.} at the level of free $\beta\gamma$ systems.

\subsection{Minimal coupling in the $\mathsf{CP}^1$ model.}\label{CP1} 
Let us consider the sum of holomorphic and anti-holomorphic copies of the supersymmetric curved $\beta\gamma$ systems with target space $\mathsf{CP}^1$. The Lagrangian of this model is (see Section \ref{GrSec})
\begin{equation}
\label{freeCP1}
\mathcal{S} = \int\mathrm{d}^2z\Big(v\cdot \thickbar{D}u+b\cdot \thickbar{D}c +i\thickbar{\mathcal{W}}(b\cdot u)-c.c.\Big)\,.
\end{equation}
Here $u$ is a two-component vector representing homogeneous coordinates on $\mathsf{CP}^1$, $v$ is a covector, the fermions $b, c$ are their superpartners respectively, and the dot denotes the ordinary scalar product (the contraction of indices). The definition of  covariant derivatives is the same as in the previous sections.

The action~(\ref{freeCP1}) is the ``free'' limit of the action~(\ref{FinalActionSGr}) for Lagrangian Grassmannians in the special case $m=1$ (up to a redefinition $v_j\rightarrow v_i\,\omega_{ij}$). Indeed, in the case $m=1$ the constraints $U^t\circ U=0$ and $C^t\circ U+U^t\circ C=0$ are identities and simply disappear from the action. In section \ref{OGrSec} we have already proven  supersymmetry of the action, but the relation of the fermions $b$ and $c$ to the tangent bundle was not immediately obvious. Our goal is to make this relation manifest.

Let us recall some relevant mathematical background\footnote{For basics on the structure of holomorphic vector bundles over complex projective spaces we refer the reader to \cite{EGUCHI1980213,okonek2013vector,GriffithsHarris}.}. Recall that $\mathscr{O}(n)$ line bundles over $\mathsf{CP}^1$ are defined by the patching function $g_{12} = \displaystyle \bigg(\frac{u_2}{u_1}\bigg)^n$ between two standard charts $\{u_1\neq 0\}$ and $\{u_2\neq 0\}$. For $n>0$, this bundle has holomorphic sections that  can be identified with degree $n$ homogeneous polynomials in two variables \big(the homogeneous coordinates on $\mathsf{CP}^1$\big). 
The standard result from complex algebraic geometry says that the tangent bundle $\mathrm{T}\mathsf{CP}^1$ is naturally isomorphic to the $\mathscr{O}(2)$ bundle. 
Let us recall the idea of the proof. Again, consider the   chart with $u_1\neq 0$ and local coordinate $z={u_2\over u_1}$. The vector field $\partial_z$ is a section of the tangent bundle in this chart. Suppose $w={u_1\over u_2}$ is a coordinate in the second chart, so that $w=z^{-1}$ in the intersection. Then $\partial_z=-w^2\partial_w$. Thus, the section of the tangent bundle transforms precisely as a homogeneous polynomial of degree $2$, \text{i.e.} as a section of $\mathscr{O}(2)$ (the extra minus can be removed by the sign swap $w\to -w$). 
In this notation the tautological bundle $\mathcal{S}$ is~$\mathscr{O}(-1)$, so that
\begin{gather}
\mathrm{T}\mathsf{CP}^1=\mathscr{O}(2)\simeq\big(\mathcal{S}^\vee\big)^{\otimes 2}\,,
\end{gather}
where $\mathcal{S}^\vee$ is the dual of the tautological line bundle. This is in accord with the general formula~(\ref{tanbun}) in the special case $m=1$.

Let us find a formulation of the model~(\ref{freeCP1}) where the $\mathscr{O}(2)$ bundle would be manifest. First, we can solve the constraint $b\cdot u = 0$ as 
\begin{gather}\label{biresol}
    b_i = h\,\omega_{ij}u_j\,,
\end{gather}
where the indices $i,j\in\{1,2\}$ label components of the (co)vectors, $\omega_{ij}$ represents the standard skew-symmetric matrix $\omega = \begin{pmatrix}
0 & 1\\
-1 & 0
\end{pmatrix}$, whereas $h$ is a new fermionic field.  Let us introduce another fermionic field by means of the formula
\begin{gather}
    f=c_i\,\omega_{ij}\,u_j\,.
\end{gather}
Under the gauge transformation  $u_i\rightarrow gu_i$, where $g\in\mathbb{C}^\ast$, our fields transform as 
\begin{align}
v_i\rightarrow g^{-1}v_i,\qquad c_i\rightarrow gc_i,\qquad b_i\rightarrow g^{-1}b_i,\qquad f\rightarrow g^2f,\qquad h\rightarrow g^{-2}h.
\end{align}
Note that the field $f$ transforms exactly as a section of $\mathscr{O}(2)$. We can therefore conclude that the  fields in the (geometric) formulation of the theory with phase space $\mathrm{T}^\ast\Pi \mathscr{O}(2)$ are precisely $\{u_i, v_i, f, h\}$. To find the formulation in terms of these fields, we substitute the expression~(\ref{biresol}) for the $b_i$ into (\ref{freeCP1}),
integrate by parts and make the change of variables $v_j\rightarrow v_j+h\,c_i\,\omega_{ij}$. After these manipulations one can rewrite the action as
\begin{align}
\mathcal{S} \simeq \int\mathrm{d}^2z \Big(v_i\thickbar{D}u_i+ h\thickbar{D}f - c.c.\Big)\,,    
\end{align}
where we have defined $\thickbar{D}f = \thickbar{\partial}f - 2i\thickbar{\mathcal{A}}f$. It is this simple action that reflects the geometric aspect of the model. Notice that the covariant derivative includes the correct charge of the field $f$ with respect to the gauge group. 

One can check that this model is also supersymmetric. However, we will not discuss this separately: in principle, this relies on the supersymmetry of the original action and the equivalence between the two theories. Instead, we will apply the ideas presented in this section to the discussion of maximal orthogonal and symplectic Grassmannians, and we shall prove supersymmetry in that more general case.

\subsection{Maximal orthogonal Grassmannian.} 
To find the reformulation of the supersymmetric sigma model with target space $\mathsf{OGr}(m,2m)$, we start with the action~(\ref{FinalActionOGr}). We will be using the same strategy as in the previous section. All arguments from this section naturally carry over to the case of Lagrangian Grassmannians. 

The key point is that in the maximal case the constraint $B\circ U=0$ may be explicitly resolved, just like in the $\CP^1$ case above. Indeed, if $U^t\circ U= 0$ and $U$ is of maximal rank~$m$, $B\circ U= 0$ implies
\begin{gather}
    B = H U^t
\end{gather}
with some $m\times m$ fermionic matrix field $H$.
Using part of the gauge symmetry (\ref{GaugeGroupOGr2}) $B\rightarrow B - y\,U^t$ (where $y$ is symmetric) one can make $H$ skew-symmetric.
We also introduce a new $m\times m$ fermionic matrix field
\begin{gather}
    F = U^t\circ C\,.
\end{gather}
Notice that the constraint $C^t\circ U+U^t\circ C = 0$ implies that $F+F^t = 0$.   
Substituting the new skew-symmetric fields $F$ and~$H$ in the action (\ref{FinalActionOGr}), performing the nonlinear shift of the bosonic field $V \rightarrow V - (CH)^t$ and integrating by parts, we get 
\begin{align}
\mathcal{S} \simeq \int\mathrm{d}^2 z\, \mathrm{Tr}\,\Big[2\big(V\circ  \thickbar{D}U+H\thickbar{D}F\big)+i\thickbar{\mathcal{R}}\,U^t\circ U-c.c.\Big]+\frac{\vkappa}{4\pi} \int\mathrm{d}^2 z\, \mathrm{Tr}\left({J}_{\mathsf{so}}\thickbar{{J}}_{\mathsf{so}}\right).\label{MinimalMaximalOGr}
\end{align}
The virtue of this formulation is that the interaction term is significantly simplified. Indeed, 
the global `supercurrent' (\ref{CurrentSO}) $\mathcal{J}_{\mathsf{so}}$ has turned into the purely bosonic one ${J}_{\mathsf{so}} = 2\pi\big(UV - V^tU^t\big)\mathds{h}_{2m}$. Thus, what we have arrived at is a bosonic Gross-Neveu model minimally coupled to  skew-symmetric fermions by the gauge field.  

The covariant derivative $\thickbar{D}F = \thickbar{\partial}F + i\thickbar{\mathcal{A}}^{\,t}F-iF\thickbar{\mathcal{A}}$ in~(\ref{MinimalMaximalOGr}) is consistent with the skew-symmetric representation of the $F$ field. The e.o.m. of the gauge field $\thickbar{\mathcal{A}}$ gives the relevant counterpart of the constraint~(\ref{constr1}):
\begin{eqnarray}
    V\circ U+2HF = 0\,.
\end{eqnarray}

\subsection{Symmetries in ``minimal formulation''.}
The gauge symmetry group in this formulation is~(\ref{bosgaugegrouportho}) (without any additional fermionic part), and the relevant  transformations
in this formulation of $\mathsf{OGr}(m,2m)$ supersymmetric sigma model (\ref{MinimalMaximalOGr}) are\footnote{The gauge where $H$ is skew-symmetric has already been fixed.}
\begin{align}
&U\rightarrow Ug\,,\quad V\rightarrow g^{-1}V+q\,U^t\,,\label{GaugeGroupMinOGr1}\\
&F\rightarrow g^tFg\,,\quad H\rightarrow g^{-1}Hg^{-1t}\,,\label{GaugeGroupMinOGr2}\\
&g\in \mathbf{GL}(m,\mathbb{C})\,,\quad q\in\mathrm{Mat}^{\text{symm}}_m\,.\label{GaugeGroupMinOGr3}
\end{align}

Supersymmetry of the model (\ref{MinimalMaximalOGr}) is not manifest, but the corresponding transformations can be derived from~(\ref{SusyActionFinal}). They have the form 
\begin{align}
\delta V = \epsilon_2(DH)U^t + 2\epsilon_2& HDU^t-\epsilon_1PV^tFP\,, \qquad\delta H = - \epsilon_1 \frac{1}{2}\big(VP^t-PV^t\big)\,,\\
&\delta U = \epsilon_1 P^tF\,,\qquad \delta F = -\epsilon_2 U^t DU\,.
\end{align}
Here, as usual, for simplicity we have set $\mathds{h}_{2m}=\mathds{1}_{2m}$. 
We leave the derivation of these transformations and the verification of the invariance of the action for Appendix~\ref{SUSYtransapp}. These transformations have the peculiarity that the  supersymmetry algebra closes up to gauge transformations.

\subsection{Lagrangian Grassmannian.} 
The representation of the supersymmetric Lagrangian Grassmannian sigma model as a model with minimally coupled fermions is parallel to the orthogonal case. By a slight modification of the arguments from the previous section, one can prove that the action (\ref{FinalActionSGr}) corresponding to $\mathsf{SGr}(m,2m)$ is equivalent to
\begin{align}
\mathcal{S} \simeq \int\mathrm{d}^2 z\, \mathrm{Tr}\,\Big[2\big(V&\circ  \thickbar{D}U+H\thickbar{D}F+i\thickbar{\mathcal{R}}\,U^t\circ U-c.c.\Big]+\frac{\vkappa}{4\pi} \int\mathrm{d}^2 z\, \mathrm{Tr}\left({J}_{\mathsf{sp}}\thickbar{{J}}_{\mathsf{sp}}\right).\label{MinimalMaximalLGr}
\end{align}
Here $H$ and $F$ are the symmetric fermionic $m\times m$ matrix fields, $\circ$ denotes contraction with a symplectic matrix $\omega_{2m}$ and ${J}_{\mathsf{sp}}$ is the bosonic part of the global supercurrent~$\mathcal{J}_{\mathsf{sp}}$.

\section{Conclusion and outlook}
In this paper we have constructed $\mathcal{N}=(2,2)$ supersymmetric generalizations of  Gross-Neveu models that correspond to  sigma models with Hermitian symmetric target spaces. We have also found that, for non-symmetric Grassmannians, there are obstacles to supersymmetry, which is consistent with the fact that the target space metrics are not K\"ahler in this case. If one tries to overcome these obstacles by imposing appropriate additional constraints, the geometric picture of the sigma model collapses (i.e. the phase space is no longer a cotangent bundle to the would-be `target space'). We also note that, in order to maintain the geometric structure of the sigma model, it is necessary that certain matrix fields have maximal possible rank. However, if we relax this requirement and impose additional conditions to preserve supersymmetry, the resulting models are self-consistent and may be of independent interest. It would also certainly be useful to develop the Gross-Neveu formulation (both bosonic and supersymmetric) for the exceptional symmetric spaces $\frac{{\mathbf E}_6}{{\mathbf{SO}}(10) \times {\mathbf U}(1)}$ and $\frac{{\mathbf E}_7}{{\mathbf E}_6 \times {\mathbf U}(1)}.$ 

Additionally, for the maximal orthogonal and symplectic (Lagrangian) Grassmannians, we have found a classically equivalent formulation that reveals the target space geometry in a simple and transparent way. Surprisingly, these supersymmetric models turn out to be purely bosonic models minimally coupled to fermions via a gauge field~$A$. 
The equivalence of the models at the quantum level might be spoiled by anomalies, and a full clarification of these matters is left for future research. In particular, it is not immediately clear how the anomaly cancellation conditions match in the two formulations. 
The novel reformulation could be useful for the study of quantum mechanical reductions. The corresponding one-dimensional quantum problems are related to Laplace and Dirac operators on the target spaces, and the minimally coupled system is likely to simplify the spectral problem for these operators.

Last but not least, another open direction is the superfield formulation of generalized supersymmetric Gross-Neveu models, which would generalize our description of free systems in section~\ref{gaugesymmsec} above.  The first steps in this matter were made in \cite{BykovPribytok} in the Abelian (Thirring) case.  
We plan to return to these questions in the future.

\vspace{1cm}
\textbf{Acknowledgments.} Sections 1-2 were written with the support of the Foundation for the Advancement of Theoretical Physics and Mathematics ``BASIS''. Sections 3-5 were supported by the Russian Science Foundation grant № 20-72-10144 (\href{https://rscf.ru/en/project/20-72-10144/}{\emph{https://rscf.ru/en/project/20-72-10144/}}). We would like to thank A.~Budekhina, A.~Fonarev, A.~Kuzovchikov, M.~Markov, A.~Smilga and F.~Vylegzhanin for discussions. DB would like to thank the Beijing Institute of Mathematical Sciences and Applications (BIMSA), where this work was finalized, for hospitality.
\vspace{1cm}

\appendix

\section{Minimal fermion coupling in $\mathsf{CP}^1$ case}\label{MinFermCP1}
In section~\ref{MinimalCoupling} we  derived supersymmetric actions with minimally coupled fermions from supersymmetrized Gross-Neveu actions. It might be more convincing, though, if one could prove that the new models are indeed equivalent to sigma models. In this section, by direct elimination of auxiliary fields, we prove that in the simplest cases $\mathsf{OGr}^+(2,4)\simeq \mathsf{CP}^1$ and $\mathsf{LGr}(1,2)\simeq \mathsf{CP}^1$ our models with minimally coupled fermions are  equivalent to the well known $\mathsf{CP}^1$ supersymmetric sigma model. The action of this model in geometric form is (see, for example, section 3 in \cite{BykovPribytok})
\begin{equation}
\label{CP1Action}
\mathcal{S}_{\mathsf{CP}^1} = \int\mathrm{d}^2z \left(g_{u\Bar{u}}\big|\,\thickbar{\partial}u\,\big|^2+ b\thickbar{\nabla} c - \bar{c}\,\nabla\bar{b} + g^{u\bar{u}}g^{u\bar{u}}R_{u\bar{u}u\bar{u}}\,bc\bar{b}\bar{c}\right)\,. 
\end{equation}
Here $ds^2=g_{u\bar{u}}\,\mathrm{d}u \mathrm{d}\bar{u}$ is the Fubini-Study metric on $\mathsf{CP}^1$ (or the round metric on the sphere, written in complex coordinates), $\nabla$ and $\thickbar{\nabla}$ are components of the Levi-Civita connection constructed from $g$, and $R_{u\bar{u}u\bar{u}}$ is the only non-vanishing component of the Riemann curvature tensor. 
The explicit formulas in convenient normalization are 
\begin{align}
&g_{u\bar{u}} = \frac{1}{2\pi\vkappa\big(1+|u|^2\big)^2}\,,\qquad g^{u\bar{u}}g^{u\bar{u}}R_{u\bar{u}u\bar{u}} = 4\pi\vkappa\,,\\
&\thickbar{\nabla} c = \thickbar{\partial} c + c\,\mathrm{\Gamma}_{uu}^{u}\thickbar{\partial} u\,,\qquad
{\nabla}\bar{b} = {\partial} \bar{b} - \bar{b}\,\mathrm{\Gamma}_{\bar{u}\bar{u}}^{\bar{u}}\partial\,\bar{u}\,,\\
&\mathrm{\Gamma}_{uu}^{u} = -\frac{2\,\bar{u}}{1+|u|^2}\,,\qquad \mathrm{\Gamma}_{\bar{u}\bar{u}}^{\bar{u}} = -\frac{2u}{1+|u|^2}\,,\label{ChristoffelOGr}
\end{align}
where $\mathrm{\Gamma}_{uu}^{u}$ and $\mathrm{\Gamma}_{\bar{u}\bar{u}}^{\bar{u}}$ are the corresponding Christoffel symbols.

\subsection{$\mathsf{LGr}(1,2)\simeq \mathsf{CP}^1$} The simplest special case of the action (\ref{MinimalMaximalLGr}) is when ${m = 1}$, which corresponds to the target space $\mathsf{LGr}(1,2)$. But there is also the obvious isomorphism $\mathsf{LGr}(1,2)\simeq\mathsf{CP}^1$, see section \ref{CP1}. Thus, it is natural to conjecture that this theory is  equivalent to the supersymmetric $\mathsf{CP}^1$ sigma model~(\ref{CP1Action}). Let us prove this. 

Parameterizing our fields as
\begin{equation}
V = \begin{pmatrix}
v_1 & v_2
\end{pmatrix}\,,\qquad
U = \begin{pmatrix}
u_1 \\ u_2
\end{pmatrix},\qquad H := b\,,\qquad F:=c
\end{equation}
and picking $\omega_2 = \begin{pmatrix}
0 & 1\\
-1 & 0
\end{pmatrix}\,,$ 
we can rewrite the action of the  $\mathsf{LGr}(1,2)$ model as
\begin{align}
\mathcal{S} =& \int\mathrm{d}^2z\,\bigg[2\Big(v_1\thickbar{\partial}u_2-v_2\thickbar{\partial}u_1+b\thickbar{\partial}c\Big)-i\thickbar{\mathcal{A}}\big(v_1u_2-v_2u_1+2bc\big)-c.c.+\nonumber\\+&2\pi\vkappa\big(|u_1|^2+|u_2|^2\big)\big(|v_1|^2+|v_2|^2\big)+2\pi\vkappa\big(u_1\bar{v}_1+u_2\bar{v}_2\big)\big(v_1\bar{u}_1+v_2\bar{u}_2\big)\bigg]\,.
\end{align}
We can use the gauge freedom $U\rightarrow Ug$ to set, say, $u_2 =1$. Eliminating $v_1$ and $v_2$ by means of the e.o.m., we get 
\begin{equation}
\mathcal{S} = 2\int\mathrm{d}^2z \bigg(\alpha\big|\,\thickbar{\partial}u\,\big|^2+\big( b\,\thickbar{\partial}\,c+\beta\, bc - c.c.\big) + \gamma \,bc\bar{b}\bar{c}\bigg)\,,
\end{equation}
where we have relabelled $u_1 := u$, and
\begin{align}
\label{CoeffLGrCp1}
\alpha = \frac{1}{2\pi\vkappa}\frac{1}{\big(1+|u|^2\big)^2}\,,\qquad\beta = -\frac{2\,\bar{u}\,\thickbar{\partial}u}{1+|u|^2}\,,\qquad \gamma = 4\pi\vkappa\,.
\end{align}
Up to an overall factor this action  coincides with (\ref{CP1Action}).

\subsection{$\mathsf{OGr}^+(2,4)\simeq \mathsf{CP}^1$}
There is the special isomorphism of complex projective manifolds\footnote{We do not consider the ``simpler'' example $\mathsf{OGr}(1,2)$ because it is a trivial variety (the disjoint union of two points).} $\mathsf{OGr}(2,4)\simeq\mathsf{CP}^1\bigsqcup\mathsf{CP}^1$. Quite generally, maximal Grassmannians $\mathsf{OGr}(m,2m)$ have two connected components\footnote{See the Appendix A.1.1. in \cite{BykovKrivorolGSM} for details.} and we denote one of these by  $\mathsf{OGr}^+(m,2m)$, thus $\mathsf{OGr}^+(2,4)\simeq\mathsf{CP}^1$. This isomorphism implies that the action (\ref{MinimalMaximalOGr}) in the case $m=2$ should be equivalent to the  supersymmetric $\mathsf{CP}^1$ sigma model. 

Let us prove that the action (\ref{CP1Action}) is equivalent (up to  normalization) to the action (\ref{MinimalMaximalOGr}) in the case of $\mathsf{OGr}^+(2,4)$. In this section, for simplicity, we set $\mathds{h}_{4}=\mathds{1}_{4}$. 
Let us discuss a parameterization of the $\{V,U,H,F\}$ fields:
\begin{itemize}
\item Using the gauge transformation  $U\rightarrow Ug$ and solving the isotropy constraints for~$U$, we get\footnote{This parameterization differs from the one used in Appendix D of~\cite{BykovKrivorolGSM} up to multiplication by the matrix $\begin{pmatrix}
2iu & 0\\
0 & -2u
\end{pmatrix}^{-1}$.
}
\begin{equation}
U = \begin{pmatrix}
1 & 0 \\
0 & 1 \\
\frac{u^2-1}{2u} & \frac{i(u^2+1)}{2u} \\
-\frac{i(u^2+1)}{2u} & \frac{u^2-1}{2u}
\end{pmatrix}\,.\label{VeroneseParameterization}
\end{equation}
Here it is implied that we have restricted to a given connected component $\mathsf{OGr}^+(2,4)$, and $u$ plays role of inhomogeneous coordinate on the sphere $\mathsf{CP}^1$. 
\item The fermion fields $F$ and $H$ are $2\times 2$ skew-symmetric matrices, thus the obvious parameterization is
\begin{equation}
\label{ParameterizationFH}
F = \begin{pmatrix}
0 & c \\
-c & 0
\end{pmatrix} = c\,\omega_2\,,\qquad
H = \begin{pmatrix}
0 & -b \\
b & 0
\end{pmatrix} = -b\,\omega_2\,.
\end{equation} 
\item Note that we have not yet used a gauge transformation of the form $V\rightarrow V+ q\,U^t$, with $q$ a  symmetric $m\times m$ matrix. Let us split the $V$ and $U$ fields into $2\times 2$ blocks as 
\begin{equation}
V = \begin{pmatrix}
V_1 & V_2    
\end{pmatrix},\qquad U = 
\begin{pmatrix}
U_1 \\ U_2
\end{pmatrix}.
\end{equation}
Using the fact that, in the parameterization (\ref{VeroneseParameterization}),  $U_1=\mathds{1}_2$ and choosing $$q = -\frac{1}{2}\left(V_1+V_1^t\right)\,,$$
 one can eliminate the symmetric part of $V_1$. Thus, one may parameterize $V$ as
\begin{equation}
\label{ParameterizationV}
V = \begin{pmatrix}
0 & v & v_1 & v_2\\
-v & 0 & v_3 & v_4  
\end{pmatrix}\,.
\end{equation}
\end{itemize}
Substituting the parameterization (\ref{VeroneseParameterization}-\ref{ParameterizationV}) into the action (\ref{MinimalMaximalOGr})
and integrating out the components of $V$, we obtain the action
\begin{align}\label{CP1OGr}
\mathcal{S} = 4\int\mathrm{d}^2z \bigg(\alpha\big|\,\thickbar{\partial}u\,\big|^2+\big( b\,\thickbar{\partial}\,c+\beta\, bc - c.c.\big) + \gamma \,bc\bar{b}\bar{c}\bigg)\,,
\end{align}
where 
\begin{align}
\label{CoeffOGrCp1}
\alpha = \frac{1}{2\pi\vkappa}\frac{1}{\big(1+|u|^2\big)^2}\,,\qquad\beta =\frac{\bar{u}\,\thickbar{\partial}u\big(1-|u|^2\big)}{|u|^2\big(1+|u|^2\big)}\,,\qquad \gamma = 4\pi\vkappa\,.
\end{align}
One sees that~(\ref{CP1OGr}) is classically equivalent to the model (\ref{CP1Action}) up to the overall factor~$4$ and the field redefinition   $c\rightarrow u^{-1}\,c$, $b\rightarrow ub$ in (\ref{CP1Action}). The latter are needed to compensate the discrepancy in the `connection part' of the fermionic $bc$ term. Indeed, using (\ref{CoeffOGrCp1}) and (\ref{ChristoffelOGr}), this discrepancy is
\begin{equation}
\beta - \mathrm{\Gamma}_{uu}^{u}\thickbar{\partial} u = \frac{\bar{u}\thickbar{\partial}u\big(1-|u|^2\big)}{|u|^2\big(1+|u|^2\big)} + \frac{2\,\bar{u}\thickbar{\partial} u}{1+|u|^2} = \frac{\thickbar{\partial}u}{u}\,.
\end{equation}
It can be eliminated by the  above field redefinition in (\ref{CP1Action}), since it effectively amounts to the substitution $b\thickbar{\partial}c\rightarrow b\thickbar{\partial}c - \big(u^{-1}\,\thickbar{\partial}u\big) bc$. 

This completes the proof that the minimally coupled supersymmetric $\mathsf{OGr}^+(2,4)$ model~(\ref{MinimalMaximalOGr}) is equivalent to the supersymmetric $\mathsf{CP^1}$ sigma model (\ref{CP1Action}).

\section{Details of supersymmetry transformations for models with
minimally coupled fermions} \label{SUSYtransapp}

The action (\ref{MinimalMaximalOGr}) was derived from the manifestly on-shell supersymmetric action (\ref{FinalActionOGr}). Therefore the action (\ref{MinimalMaximalOGr}) is also expected to be supersymmetric. In this section we shall derive the SUSY transformations of the $\{V, U, H, F\}$ fields and check supersymmetry invariance of the action. First of all, note that the number of independent components of the $V$ and $U$ fields does not equal the number of independent components of the $H$ and $F$ fields. This is due to the  gauge redundancy: if one takes it into account, the theorem on the equality of  boson and fermion degrees of freedom is seen to hold. One of the consequences of this observation is that the supersymmetry transformations of the $\{ V, U, H, F \}$ fields must necessarily contain gauge transformations.

\subsection{Derivation of the SUSY transformations.}

We shall focus on the maximal orthogonal Grassmannian, since the symplectic case may be considered with minor modifications. We set $\mathds{h}_n = \mathds{1}_{2m}$ to simplify the  calculations. To derive the supersymmetry transformations of the $\{V, U, H, F\}$ fields, we start from the transformations of the original fields $\{V, U, B, C\}$, see (\ref{SusyActionFinal}). After the shift $V \rightarrow V - HC^t$ they change slightly, so that 
\begin{equation}
\label{ShiftedSUSY}
\delta U = \epsilon_1\, C,\quad \delta C = -\epsilon_2 \,D U\,,\quad\delta (V-HC^t)=\epsilon_2\, D B,\quad \delta B = -\epsilon_1 (V-HC^t).
\end{equation}

In Section \ref{OGrSec} we explained that, for maximal Grassmannians and under the assumption that $U$ is of maximal rank, there is a useful basis in~$\mathbb{C}^n$  consisting of the columns of $U$ and $U^\ast$. Thus, we can decompose $C$  as
\begin{eqnarray}\label{Cdecomp}
    C=U\lambda + U^\ast\beta
\end{eqnarray} 
with some fermionic $m\times m$ matrices $\lambda$ and $\beta$. One can then use the gauge symmetry $C\rightarrow C+ U\,e$ \big(see (\ref{GaugeGroupOGr2})\big) to set $\lambda = 0$. 
Multiplying the decomposition of $C$ by $U^t$ we get
$\beta = \big(\thickbar{U}U\big)^{-1t}F$, where we have taken into account that $F=U^tC$. Clearly, these manipulations are equivalent to multiplying $C$ by the l.h.s. of the partition of unity~(\ref{partunity}).
Using these formulas, one can express the  transformation law of $U$ as\footnote{Recall that $P=\big(\thickbar{U}U\big)^{-1}\thickbar{U}$.}
\begin{equation}
\delta U = \epsilon_1\, C = 
\epsilon_1 P^tF\,.
\label{SUSYtransformationsOfU}
\end{equation}

To find a supersymmetry transformation of $F$, we use $\delta F = \delta(U^tC)$. Using (\ref{ShiftedSUSY}) and~(\ref{Cdecomp}), we get
\begin{align}
&\delta F =  - \epsilon_2 U^t\partial U\,.\label{SUSYfieldTransformationOfF}
\end{align}

Next, we find a transformation law for $H$.
Since $B=HU^t$, one finds formally 
$\delta H = -\epsilon_1 VP^t\,.$
In general this is not skew-symmetric and thus in conflict with our assumption that $H$ is a skew-symmetric matrix. The paradox may be resolved by applying an additional gauge transformation $H\rightarrow H+ y$, with $ y$ a symmetric fermionic matrix (see~(\ref{GaugeGroupOGr2})). 
Choosing for $y$ (minus) the symmetric part of the above variation~$\delta H$,  
\begin{equation}\label{y}
y= -\frac{1}{2}\Big(\epsilon_1 VP^t+\big(\epsilon_1 VP^t\big)^t\Big)\,,
\end{equation}
we bring $H$ back to skew-symmetric form. 
Altogether the supersymmetry transformation is cast in the form
\begin{align}
&\delta H =
y -\frac{1}{2}\epsilon_1\big(VP^t-PV^t\big)\,,\label{SUSYTransformationOfH}
\end{align}
where $y$ is given by~(\ref{y}).

Deriving an analogous formula for  $V$  is a more technical exercise, essentially because $V$ has the most complicated gauge transformation structure, see~(\ref{GaugeGroupOGr1}). Here one should use the partition of unity, as well as the constraint $VU+2HF = 0$. In order not to clutter the presentation, we give only the final answer:
\begin{align}
&\delta V = 
 \delta_{y} V -\epsilon_1 PV^tFP + \epsilon_2\Big((DH)U^t+2H(DU)^t\Big)\,,\qquad \delta_y V := -2yFP\,.\label{SUSYTransformationOfV}
\end{align}
Here, as before, $y$ is given by  formula (\ref{y}).

Thus, the supersymmetry transformations for the model with minimally coupled fermions are given by  (\ref{SUSYtransformationsOfU}, \ref{SUSYfieldTransformationOfF}, \ref{SUSYTransformationOfH}, \ref{SUSYTransformationOfV}). In these formulas the additional gauge transformations $\delta_y H:=y$ and $\delta_y V$ have been written out explicitly to explain the derivation from the supersymmetry transformations of the original model; at the end these gauge transformations may be safely dropped.

\subsection{SUSY invariance of the action.}
In section~\ref{MinimalCoupling} we have derived actions that reflect geometric information about Gross-Neveu models with fermions. However, the resulting supersymmetry transformations are more complicated. 
In this section we shall check  invariance of the action (\ref{MinimalMaximalOGr}) corresponding to maximal orthogonal Grassmannians w.r.t. the supersymmetry transformations\footnote{The original SUSY transformations derived in the previous section involved some additional redundant  gauge transformations. Here we omit these gauge transformations.} (\ref{SUSYtransformationsOfU}, 
\ref{SUSYfieldTransformationOfF}, \ref{SUSYTransformationOfH},  \ref{SUSYTransformationOfV}) derived in the previous section:
\begin{align}
&\delta V = \epsilon_2(DH)U^t + 2\epsilon_2 HDU^t-\epsilon_1PV^tFP\,, \label{deltaV}\qquad\delta H = - \epsilon_1 \frac{1}{2}\big(VP^t-PV^t\big)\,,\\
&\delta U = \epsilon_1 P^tF\,,\qquad \delta F = -\epsilon_2 U^t DU\,.\label{deltaF}
\end{align}

Let us start with the free (gauged) holomorphic part of the action
\begin{equation}
\mathcal{S}_0 = \int\mathrm{d}^2z\,\mathrm{Tr}\Big(V\thickbar{D}U + H\thickbar{D}F\Big)\,.
\end{equation}
Variation of this action w.r.t. the  transformations (\ref{deltaV}-\ref{deltaF}) has the form
\begin{align}
\delta\mathcal{S}_0 = \,\epsilon_1\int\mathrm{d}^2&z\,\mathrm{Tr}\Big(V\thickbar{D}\big(P^tF\big) - PV^tFP\thickbar{D}U+PV^t\thickbar{D}F \Big) +\nonumber\\
&+\epsilon_2\int\mathrm{d}^2z\,\mathrm{Tr}\Big((DH)U^t\thickbar{D}U+2HDU^t\thickbar{D}U + H\thickbar{D}\big(U^tDU\big)\Big)\,.
\end{align}

Consider first the term proportional to $\epsilon_2$. Integrating by parts in the first term in brackets and using the fact that $U^t\big[\thickbar{D}, D\big]U\sim U^tU F_{z\bar{z}}\simeq 0$, 
one casts the term proportional to $\epsilon_2$ in the form $HDU^t\thickbar{D}U + H\thickbar{D}U^tDU$. This term is traceless due to the skew-symmetry of $H$.

Next, we consider the term proportional to $\epsilon_1$. Here it is useful to keep is mind the partition of unity~(\ref{partunity}) $UP+(UP)^t = \mathds{1}_{2m}$, the relations $PU = \mathds{1}_m$, $PP^t \simeq 0$ and their derivatives. 
Using these relations and the constraint $VU+2HF = 0$, one can simplify the term proportional to $\epsilon_1$ as follows:
\begin{align}
\mathrm{Tr}\Big(V\thickbar{D}\big(P^tF\big) - PV^tFP\thickbar{D}U+PV^t\thickbar{D}F \Big) \simeq\mathrm{Tr}\Big(2\big(\thickbar{D}P\big)P^t(FHF)\Big)\,.
\end{align}
The combination $FHF$ is symmetric, whereas  $\big(\thickbar{D}P\big)P^t$ is skew-symmetric due to $PP^t \simeq 0$. Thus, the term proportional to $\epsilon_1$   vanishes as well, and the free part of the action is invariant under SUSY transformations.   

Next, let us check the invariance of the interaction part of the action 
\begin{equation}
\mathcal{S}_{int} = \int\mathrm{d}^2 z\, \mathrm{Tr}\left({J}_{\mathsf{so}}\thickbar{{J}}_{\mathsf{so}}\right)\,.
\end{equation}
We will use the same logic as in sections \ref{GrSec}-\ref{OGrSec} above. The variation of the current is given by
\begin{equation}
\delta {J}_{\mathsf{so}} = 2\epsilon_2\partial\big(UHU^t\big)\,.
\end{equation}
Here we have used the partition of unity $UP+(UP)^t=\mathds{1}_{2m}$ to prove that the terms proportional to $\epsilon_1$ vanish.
As in section \ref{OGrSec}, we compute the commutator of $\tilde{{J}_{\mathsf{so}}}:=UHU^t$ with the current ${J}_{\mathsf{so}}$, finding  that it is proportional to the constraints and thus also vanishes on-shell.
Thus, the interaction term is also invariant under supersymmetry transformations. 

Formally, we also should check that our transformations (\ref{deltaV}-\ref{deltaF}) really constitute a proper supersymmetry algebra. If we write our supersymmetry transformations as $\delta = \epsilon_1 Q_1 +\epsilon_2 Q_2$,
then the algebra relations
\begin{eqnarray}
    \{Q_1,Q_2\} = \partial \quad\quad \textrm{and}\quad\quad Q^2_1=Q_2^2=0
\end{eqnarray}
must hold. This check is merely a technical exercise that involves repeated use of the e.o.m.  and the completeness identity. We leave this verification to the interested reader. Similarly, there are analogous calculations for the Lagrangian Grassmannians. The only changes that need to be implemented in those cases amount to restoring all instances of the $\omega_{2m}$ matrix.

\section{Tangent bundle of Grassmannians} \label{TungentBundles}
As discussed in the text, fermions in supersymmetric sigma models (i.e., in our setting this is when the target space is a quadric or a maximal orthogonal/symplectic Grassmannian)  take values in the tangent bundle of the corresponding Grasmannian\footnote{With shifted fermionic degree meaning that the fibers are assumed `fermionic'.}. Since the cases of orthogonal and symplectic cases are parallel, we consider only the orthogonal one. Here fermions can be considered as sections of $\mathsf{\Pi}\big(\mathrm{T}\mathsf{OGr}(m,n)\big)$ (for $m=1$ or ${n=2m}$). Thus it is interesting to understand the structure of the tangent bundle. 

A standard result about tangent bundles of orthogonal Grassmannians (valid for any values of $m, n$) is that there exists a short exact sequence of holomorphic vector bundles 
\begin{equation}
\label{ExactSequenseOGr}
0\longrightarrow\Big(\mathcal{S}^\vee\otimes\left(\mathcal{Q}^\vee\big/\mathcal{S}\right)\Big)\bigg|_{\mathsf{OGr}(m,n)}\longrightarrow\mathrm{T}\mathsf{OGr}(m,n)\longrightarrow\left(\bigwedge^2\mathcal{S}^\vee\right)\bigg|_{\mathsf{OGr}(m,n)}\longrightarrow 0\,.
\end{equation}
Here $\mathcal{S}$ is the tautological bundle over the Grassmannian, $\mathcal{S}^\vee$ is its dual and $\mathcal{Q}\simeq\mathbb{C}^n\big/\mathcal{S}$ is the quotient bundle. 

\subsection{``Differential-geometric'' argument.} 
Let us prove this statement in a somewhat differential-geometric manner. Consider the vector $v:=X{\dd\over \dd U}$, which is formally a section of $\mathcal{S}^\vee\otimes \mathbb{C}^n$. It is tangent to $\mathsf{OGr}(m,n)$ if
\bea\label{tanvecdef}
0=v\big(U^tU\big)=X^tU+U^tX\,.
\eea
Thus $v$ is a section of $\mathrm{T}\mathsf{OGr}(m,n)$ if and only if~(\ref{tanvecdef}) holds. On the other hand $v$ is a section of $\Big(\mathcal{S}^\vee\otimes\left(\mathcal{Q}^\vee\big/\mathcal{S}\right)\Big)\bigg|_{\mathsf{OGr}(m,n)}$ if 
\bea\label{quotbundle}
X^tU=0\quad\quad \textrm{and}\quad\quad  X\sim X+U \lambda\,,\quad\lambda \in \mathrm{Mat}_m\,.
\eea
Indeed, the first equality means that $v$ vanishes on $\mathcal{S}$ and, as such, is a section of $\mathcal{S}^\vee\otimes \mathcal{Q}^\vee$. The second identification, in turn, means that we pass to the quotient bundle $\mathcal{S}^\vee\otimes\left(\mathcal{Q}^\vee\big/\mathcal{S}\right)$.

Clearly, $X^tU=0$ implies~(\ref{tanvecdef}), which therefore proves the embedding part of~(\ref{ExactSequenseOGr}). In the second step, starting from $v$ satisfying~(\ref{tanvecdef}) one can define an element of $\Big(\bigwedge^2\mathcal{S}^\vee\Big)\bigg|_{\mathsf{OGr}(m,n)}$ by mapping
\bea\label{vYmap}
v\mapsto Y:=X^tU\,.
\eea
The r.h.s. is skew-symmetric as a consequence of~(\ref{tanvecdef}). Moreover, by the definition~(\ref{quotbundle}), the kernel of this map is precisely given by sections of ${\Big(\mathcal{S}^\vee\otimes\left(\mathcal{Q}^\vee\big/\mathcal{S}\right)\Big)\bigg|_{\mathsf{OGr}(m,n)}}$. Thus, exactness is proven. The above map is also surjective, since for any section $Y$ of $\bigwedge^2\mathcal{S}^\vee$ one can construct the respective vector field $X=-P^tY$ that is mapped to $Y$ under~(\ref{vYmap}).

\subsection{Extremal Grassmannians.} 
The sequence~(\ref{ExactSequenseOGr}) automatically determines the structure of the tangent bundles in the maximal and minimal cases,   
\begin{equation}
\label{ExtremeIsom}
\mathrm{T}\mathsf{OGr}(1,n)\simeq\Big(\mathcal{S}^\vee\otimes\left(\mathcal{Q}^\vee\big/\mathcal{S}\right)\Big)\bigg|_{\mathsf{OGr}(1,n)},
\quad\mathrm{T}\mathsf{OGr}(m,2m)\simeq\left(\bigwedge^2\mathcal{S}^\vee\right)\bigg|_{\mathsf{OGr}(m,2m)},
\end{equation}
because in each of these two cases either the first or last term in the sequence~(\ref{ExactSequenseOGr}) vanishes. For example, when $m=1$, the exterior square $\bigwedge^2\mathcal{S}^\vee=0$ by dimensional reasons. Alternatively, one has $X^tU+U^tX=2 X^tU$, so that the conditions determining the first two bundles in~(\ref{ExactSequenseOGr}) coincide. In the opposite case $n=2m$ equation $X^tU=0$ leads to $X=U\lambda$, where $\lambda\in \mathrm{Mat}_m$. This implies that $X\sim 0$ in the quotient~(\ref{quotbundle}), so that in this case the first term in~(\ref{ExactSequenseOGr}) vanishes.

The only difference in the symplectic case is that the exterior square is replaced by the symmetric one, i.e.
\begin{equation}
\mathrm{T}\mathsf{LGr}(m,2m)\simeq \left(\mathrm{Sym}^2\mathcal{S}^\vee\right)\Big|_{\mathsf{LGr}(m,2m)}\,.
\end{equation}
The difference comes from the fact that here  the $m$-planes in $\mathbb{C}^n$ are isotropic with respect to a skew-symmetric bilinear form.

\subsection{``Formal'' argument.} 
Let us mention a more formal way of proving~(\ref{tanbun}) (which, as already explained, is a special case of~(\ref{ExactSequenseOGr})). Consider, for a change, the Lagrangian Grassmannian $\mathsf{LGr}(m,2m)\hookrightarrow\mathsf{Gr}(m,2m)$ defined by a symplectic form $\omega$ on~$\mathsf{C}^n$. The ``normal exact sequence'' with respect to this embedding has the form
\begin{equation}
\label{normalsequence}
0\longrightarrow \mathrm{T}\mathsf{LGr}(m,2m)\longrightarrow\mathrm{T}\mathsf{Gr}(m,2m)\Big|_{\mathsf{LGr}(m,2m)}\longrightarrow\mathcal{N}_{\mathsf{LGr}(m,2m)}\longrightarrow 0\,.
\end{equation}
This is simply a rephrasing of the definition of normal bundle $\mathcal{N}_{\mathsf{LGr}(m,2m)}$. 
One can interpret the symplectic form as a section of  $\bigwedge^2\mathcal{S}^\vee$  and the Lagrangian Grassmannian as the vanishing locus of this section. Thus, the normal bundle is
\begin{equation}
\mathcal{N}_{\mathsf{LGr}(m,2m)}\simeq \left(\bigwedge^2\mathcal{S}^\vee\right)\bigg|_{\mathsf{LGr}(m,2m)}\,. 
\end{equation}
Besides, there is a well-known isomorphism for the tangent bundle of usual Grassmannian~\cite{eisenbud_harris_2016}, namely $\mathrm{T}\mathsf{Gr}(m,2m)\simeq \mathcal{S}^\vee\otimes\mathcal{Q}$. Substituting these facts into (\ref{normalsequence}), we obtain a short exact sequence 
\begin{equation}
\label{mainexactsequence}
0\longrightarrow \mathrm{T}\mathsf{LGr}(m,2m)\xlongrightarrow[]{\iota}\Big(\mathcal{S}^\vee\otimes\mathcal{Q}\Big)\Big|_{\mathsf{LGr}(m,2m)}\xlongrightarrow[]{\rho}\left(\bigwedge^2\mathcal{S}^\vee\right)\bigg|_{\mathsf{LGr}(m,2m)}\longrightarrow 0\,.
\end{equation}

Consider  the bundle $\mathcal{S}^\bot$ of planes in $\CC^n$ orthogonal to $\mathcal{S}$ w.r.t. the symplectic form~$\omega$. Then, $\omega$ induces an isomorphism between $\mathcal{S}$ and $\left(\mathbb{C}^n\big/\mathcal{S}^\bot\right)^\vee$. However, in the maximal (Lagrangian) case one has $\mathcal{S}\simeq\mathcal{S}^\bot$.  
Recalling that $\mathcal{Q}\simeq\mathbb{C}^n\big/\mathcal{S}$, one thus has $\mathcal{Q}\simeq \mathcal{S}^\vee$ in the maximal case.

It follows that the middle term in (\ref{mainexactsequence}) is just $\mathcal{S}^\vee\otimes\mathcal{S}^\vee\simeq\mathrm{Sym}^2\mathcal{S}^\vee\oplus\bigwedge^2\mathcal{S}^\vee$. Thus the epimorphism $\rho$ is just the projection onto the second summand, whereas $\iota$ is an identity map to the first summand. This proves the statement in the symplectic case;  the same argument  works in the maximal orthogonal case, with obvious adjustments.

There is also a formal way of proving~(\ref{ExactSequenseOGr}) for arbitrary orthogonal or symplectic Grassmannians, but the general proof is more technical. It relies on a combination of the ``normal exact sequence'' (\ref{mainexactsequence}), the ``tautological exact sequence'' $$0\longrightarrow\mathcal{S}\longrightarrow\mathbb{C}^n\longrightarrow\mathcal{Q}\longrightarrow 0$$ and the snake lemma.

\setstretch{0.8}
\setlength\bibitemsep{5pt}
\printbibliography
\end{document}